\documentclass[aps,prb,twocolumn,reprint,groupedaddress,showpacs]{revtex4-1}
\usepackage{graphicx}
\usepackage{subcaption}
\usepackage{verbatim}
\usepackage{amssymb,amsmath}

\newcommand{\ket}[1]{| #1\rangle}
\newcommand{\bra}[1]{\langle #1 |}

\newcommand{\ex}[1]{\langle #1 \rangle}

\begin{document}


\title{Conditioned outputs, distribution of decision times and measurement-based feedback scheme for continuous weak linear measurement of a simple quantum system.}


\author{A. Franquet}
\email[]{A.FranquetGonzalez@tudelft.nl}
\affiliation{Kavli Institute of Nanoscience, Delft University of Technology, 2628 CJ Delft, The Netherlands}

\author{N. C. Kruse}
\affiliation{Kavli Institute of Nanoscience, Delft University of Technology, 2628 CJ Delft, The Netherlands}

\author{B. Vervliet}
\affiliation{Kavli Institute of Nanoscience, Delft University of Technology, 2628 CJ Delft, The Netherlands}

\author{Yuli V. Nazarov}
\affiliation{Kavli Institute of Nanoscience, Delft University of Technology, 2628 CJ Delft, The Netherlands}


\date{\today}

\begin{abstract}
We address the peculiarities of the quantum measurement process in the course of a continuous weak linear measurement (CWLM). As a tool, we implement an efficient numerical simulation scheme that allows us to generate single quantum trajectories of the measured system state as well as the recorded detector signal, and study statistics of these trajectories with and without post-selection. In this scheme, a linear detector is modelled with a qubit that is weakly coupled to the quantum system measured and is subject to projective measurement and re-initialization after a time interval at each simulation step. We explain the conditions under which the scheme provides an accurate description of CWLM.\\

We restrict ourselves to a simple but generic situation of a qubit non-demolition measurement. The qubit is initially in an equal-weight superposition of two quantum states. In the course of time, the detector signal is accumulated and the superposition is destroyed. It is known that the times required to resolve the quantum states and to destroy the superposition are of the same order. We prove numerically a rather counter intuitive fact: the average detector output conditioned on the final state does not depend on time. It looks like from the very beginning, the qubit knows in which state it is. We study statistics of decision times where the decision time is defined as time required for the density matrix along a certain trajectory to reach a threshold where it is close to one of the resulting states. This statistics is useful to estimate how fast a decisive CWLM can be. \\
Basing on this, we devise and study a simple feedback scheme that attempts to keep the qubit in the equal-weight superposition. The detector readings are used to decide in which state the qubit is and which correction rotation to apply to bring it back to the superposition. We show how to optimize the feedback parameters and move towards more efficient feedback schemes.
\end{abstract}

\maketitle

\section{Introduction}

The standard description of quantum mechanics introduces projective measurement as an instantaneous non-unitary process by which a quantum system is projected into an eigenstate of a measured observable with a probability given by Born's rule. In reality, the measurements are never instantaneous but occur over some time scale that is determined by the details of the interaction between the measured system and its environment and required to obtain a reliable measurement result.
This idea has become one of the basis and principal ingredients in the study of quantum control at the core of quantum computing and communication\cite{NielsenChuang}. \\
A more general and adequate description of the measurement process is provided by the paradigm of continuous weak linear measurement (CWLM)~\cite{CWLM0,CWLM1,CWLM2,CWLM25,NazWei,CWLM3,CWLM4}.\\
Recent technological advances have made possible  to utilize and study CWLM in every detail for a set of quantum device setups.  Experiments  realize continuous measurement and monitoring of quantum systems, and even provide the information about  single quantum trajectories~\cite{Devoret, SiddiqiSingle, SiddiqiEntanglement,Huard, DiCarlo, SiddiqiMapping, SiddiqiMolmer}. This resulted in  a more elaborative  and practical understanding of the measurement process in quantum mechanics.\\
In particular, the experimental realization of interesting phenomena related to the conditioning of a quantum system using measurement and feedback is of relevance to our work~\cite{WisemanWeakValues, FranquetNaz, FranquetNaz2, SiddiqiFeedback, AnalogFeedback}.\\

In this work, we study the CWLM  implementing numerically an iterative simulation procedure that is essentially equivalent to those commonly used \cite{simulation1,simulation2} but formulated in more transparent and basic terms. This tool permits a deep investigation of the measurement process that is not possible analytically. With this, we can directly simulate individual quantum trajectories from the first principle quantum state evolution and quickly accumulate sufficiently big statistics of these trajectories to compute the distribution of various quantities characterizing the measurement, including the conditioning of the trajectories.

In contrast to usual descriptions of CWLM that are based on a Bloch equation for the density matrix of the measured system , or on stochastic differential equations, the tool gives insight not only into the characteristics of the measured system but also into the generation of a measurement signal in a linear measurement setup.
The tool is quite simple. The detector is represented by a qubit. At each step of the simulation, the qubit is first initialized to an equal-weight superposition of two states. Then for a time interval of $\Delta t$ it is coupled to the system measured. We evaluate the unitary evolution of the system and the qubit on this interval. After that, the qubit is  measured projectively, the measurement result counts for the detector output at this time interval, and the density matrix of the system is updated according to the measurement result. We show that this setup accurately reproduces CWLM at proper choice of measurement strength and the duration of the time interval.

Although the tool permits accurate simulation of rather complex quantum systems and measurement setups, in this Article we apply it to the simulation of one of the simplest yet generic situations of CWLM: the non-demolition measurement (see e.g. \cite{nondemolition}). The quantum system is a qubit. It is initially prepared in an equal-weight superposition of two quantum states, $(|+\rangle + |-\rangle)/\sqrt{2}$. It is measured in the basis of these two states. As a result of the decoherence induced by the measurements, the superposition is destroyed at certain time scale, and the density matrix of the qubit becomes diagonal. The qubit is in either of the two states. The mean value of the detector output $V(t)$ freezes at one of the two levels corresponding to the states. We normalize the signal such that these levels correspond to $v=\pm 1$. A repetitive measurement would reproduce the same result. Since the detector signal is noisy, it takes a finite time to resolve these two levels of the signal. This acquisition time is of the same order as decoherence time\cite{QNoise}. Owing to simplicity of the system, we can compare some results of the simulation with the analytical results.

We start our study with computing the average value of the detector output. Owing to symmetric initial conditions, this value is always zero. However, we can condition the output at its asymptotic mean value computing $\langle v(t) v(\infty)\rangle$. An intuitive expectation is that this quantity is 0 at $t=0$ (since the qubit is in an equal-weight superposition) and saturates at $1$ if $t\to \infty$. However, we show that the conditioned output does not depend on time. It looks like the qubit "knows" from the very beginning in which of the two states it is and the superposition is indistinguishable from an equal-weight diagonal density matrix. We confirm this counter intuitive result analytically.

In reality, an observer can not instantly decide in which state the qubit is. Let us assume that the observer has a full information about the measurement results of the detector qubit and can therefore access the density matrix of the measured qubit along the quantum trajectory at any given moment of time. He monitors the probability to be in one of the states, say, $p_+$, and waits till it achieves certain small threshold $h$. If $p_+ = 1-h/2$, he decides that the qubit is in '+' state, if $p_+ = h/2$, the qubit must be in opposite state. This moment we call {\it decision time}. This time varies from trajectory to trajectory, and we are interested in the distribution of the decision times in dependence on the threshold $h$.
This quantifies how fast the measurement can bring certain result and helps in planning an actual fast measurement.

We go into details of decision dynamics and consider the situation when the decisions are used for a feedback. As a simple example, we formulate and simulate a feedback scheme that has a purpose to keep the qubit in the equal-weight superposition. The observer accumulates the detector output during a time interval $T_f$. If the average value of the output exceeds a certain threshold, $|v|>I$, he decides the qubit is in the state ${\rm sgn}(v)$ and applies a correcting unitary transform that brings the qubit back to the equal-weight superposition.
We made detailed simulations of the feedback dynamics and attempt to optimize the average probability to be in the superposition with respect to parameters $I, T_f$. We compare the results with some analytical predictions.

The structure of the Article is as follows. We explain and present the simulation tool used in Section~\ref{model} and formulate the general description of the scheme.
In subsection~\ref{linear} we specify to the case when the detector qubit can be effectively considered as a linear detector that measures another qubit and discuss the conditions for this and the details of numerical implementation.\\
Further, in Section~\ref{results} we present the simulation results concerning the average conditioned detector output and the distribution of the decision times.
\\
In Section~\ref{sec:feedback}, we present and discuss the feedback scheme described. The subsection \ref{sec:analyticalfeedback} elaborates on the scheme on analytical level. We present the simulation results of the feedback dynamics in subsection~\ref{sec:numfeedback} and show how optimize the feedback efficiency as a function of two parameters.\\
We conclude in Section~\ref{conclusion}.

\section{The simulation tool}
\label{model}

Our goal is to describe in general a continuous measurement process using a discrete stochastic update approach. We outline a step-by-step stochastic process that will mimic a random time-line of an actual continuous measurement performed in an experimental setup.\\
Let us consider a general measurement scenario in which a quantum system A is being measured with making use of another quantum system B (the detector). The dynamics of these systems are governed  by the corresponding Hamiltonians $\hat{H}_{A}$, $\hat{H}_{B}$.\\
For the information transfer from the system measured to the detector, there must be an  interaction between those systems, a coupling of a kind between the degrees of freedom of A and B. Thus, the complete dynamics in this simple yet general scenario is governed by a total Hamiltonian:
\begin{equation}
\hat{H} = \hat{H}_{A}+\hat{H}_{B}+\hat{H}_{c},
\end{equation}
where $\hat{H}_{c}$ is the coupling Hamiltonian. For a simplest case when the detector is sensitive to a single observable $M$, the coupling Hamiltonian can be represented as $\hat{H}_{c}=\hat{M}\otimes\hat{Q}$ where $\hat{M}$ is an operator acting in system A and $\hat{Q}$ is an operator acting in B.\\
The stochastic update process we construct is supposed to simulate  the time-line of an actual experimental run where the random outputs of the detectors in short time intervals are measured  and recorded. With this in mind, the coupling at each step persists during a time interval $\Delta t$. To simulate a continuous measurement, the  $\Delta t$ should by chosen such that the change of the density matrix of the measured system $\propto \Delta t$ is small. In this limit,  the simulation process can be described with a quasi-continuous stochastic differential equation.\\

At the beginning of the simulation step, the interaction has not been switched on. The measured system and the detector are in a product state $\hat{\rho}_i=\hat{\rho}^{A}(0)\otimes\hat{\rho}^{B}(0)$. It is convenient to initialize the detector to the same $\hat{\rho}^B$ at each step. Then the whole density matrix undergoes a unitary evolution determined by $\hat{H}$. It is convenient to disregard $\hat{H}_{A}$ and $\hat{H}_{B}$ for this evolution. One can formally do this, for instance, by applying a unitary transformation that switches to the interaction picture and to disregard subsequently the time dependence of $H_c(t)$ during a short time interval $\Delta t$. Alternatively, one can separate the evolution governed by $\hat{H}_A+\hat{H}_B$ and $H_c$ in time, adding an extra simulation step of the same duration where the dynamics is governed by $\hat{H}_A+\hat{H}_B$. This is valid in the limit of small $\Delta t$ where $\exp\left(i \hat{H} \Delta t \right) \approx \exp\left(i (\hat{H}_A+\hat{H}_B)\Delta t \right) \exp\left(i \hat{H}_c \Delta t \right)$
 With this, the whole density matrix in the end of the time interval becomes

\begin{equation}
\hat{\rho}(\Delta t)=e^{-i\Delta t\hat{M}\hat{Q}}\hat{\rho}_ie^{+i\Delta t\hat{M}\hat{Q}}.
\end{equation}
\\
 One can use the eigenbasis $\ket{n}$ of the operator $\hat{M}$, $\hat{M}\ket{n}=M_{n}\ket{n}$ to rewrite the previous equation

 \begin{equation}
 \label{eq:evolution}
 \hat{\rho}(\Delta t)= \sum_{n,m}\rho^{A}_{n,m}(0)\ket{n}\bra{m}\otimes\hat{K}_{n,m}(\Delta t),
 \end{equation}

where $\hat{K}_{n,m}(\Delta t) = e^{-i\Delta t M_{n}\hat{Q}}\hat{\rho}^{B}(0)e^{+i\Delta t M_{m}\hat{Q}}$.\\
After the time interval, the detector system is projectively measured in the basis $\ket{i}$ that {\it does not} coincide with the eigenbasis of $\hat{Q}$. The probability of the outcome $i$ is given by

\begin{equation}
\label{eq:probability}
P(i) = \rm{Tr}_{A}\bra{i}\hat{\rho}(\Delta t)\ket{i}=\sum_n\rho^{A}_{n,n}(0)\bra{i}\hat{K}_{n,n}(\Delta t)\ket{i}.
\end{equation}
Here, $\rm{Tr}_{A}$ is a partial trace over the space of the system A
Once the detector is projected to  the  state $i$, and the result is recorded, the density matrix of the system measured becomes

\begin{equation}
\label{eq:update}
\rho^{A}_{new}(\Delta t) = \frac{\sum_{n,m}\rho^{A}_{n,m}(0)\ket{n}\bra{m}\bra{i}\hat{K}_{n,m}(\Delta t)\ket{i}}{\sum_n\rho^{A}_{n,n}(0)\bra{i}\hat{K}_{n,n}(\Delta t)\ket{i}}.
\end{equation}

This density matrix is taken as the initial one $\hat{\rho}_A$ at the next step of the simulation. The detector is initialized again to $\hat{\rho}^{B}(0)$  and the step is repeated.\\
With this procedure, the random outputs of the detector are recorded like eventual readings in an experiment while the measured system undergoes a stochastic update process. The random outputs of the detector can then be combined in a random time-dependent variable $V(t)$ which due to the previous derivation will contain information about the measured system expected values of the operator $\hat{M}$.  As discussed in the following section, this simulates CWLM provided the strength of the interaction at each step ($\hat{M}\Delta t$) is small.\\
\\

While any quantum system is in principle suitable to simulate a detector, here we concentrate on a simplest one and consider a qubit.

\subsection{Qubit as a linear detector}
\label{linear}

Let us consider a qubit that measures an operator $\hat{M}$ in the space of the system A. In general, this operator may be associated with an effective magnetic field acting on the qubit pseudo-spin. This magnetic field causes precession of the pseudo-spin with the angle directly proportional to this magnetic field. This leads to a straightforward setup of an approximately linear qubit detector. Initially, the qubit pseudo-spin is in $x$ direction. Let the magnetic field rotate it in $y$ direction. This will cause the deviation of the pseudo-spin in $z$ direction that is linear in $\hat{M}$ in the limit of small $\hat{M} \Delta t$, this deviation being detected in \\

To quantify, we note that initially the whole system is a product state $\hat{\rho}(0)=\hat{\rho}^{A}\otimes\ket{x}\bra{x}$ (where $\hat{\sigma}_x\ket{x}=\ket{x}$ is an eigenstate of the Pauli matrix $\hat{\sigma}_x$).  At a step, we turn on the coupling Hamiltonian $\hat{H}_c=\hat{M}\otimes\hat{\sigma}_y$ for the duration $\Delta t$ of the step. By the end of the step,
the resulting density matrix in the eigenbasis of the operator $\hat{M}$ reads
\begin{align}
\label{eq:evolution2}
\hat{\rho}_{n,m}(\Delta t) &= \hat{\rho}^{A}_{n,m}(0)\otimes e^{-iM_{n}\hat{\sigma}_y\Delta t}\ket{x}\bra{x}e^{ iM_{m}\hat{\sigma}_y\Delta t}\\ \nonumber
 &= \hat{\rho}^{A}_{n,m}(0)\otimes\hat{K}_{n,m}(\Delta t),
\end{align}

$M_{n}$ being the eigenvalues $\hat{M}$.

In the end of the detector qubit is projected onto the Z basis and a result of $\pm 1$ is recorded, with the probability given by Eq. \eqref{eq:probability},
\begin{equation}
\label{eq:probability2}
P(\pm) =\frac{1}{2}\sum_n\rho^{A}_{n,n}(0)\left(1\pm\sin(2M_n\Delta t)\right)
\end{equation}
We see that $\langle \hat{\sigma}_z \rangle = 2 \Delta t \langle \hat{M}\rangle$ in the limit of $\hat{M} \Delta t \to 0$, as one expects from the linear measurement.
\\
Finally, the density matrix of the system A  is updated depending on the detector reading $\pm1$ according to Eq. \eqref{eq:update},
\begin{subequations}
\begin{align}
\label{eq:update2}
\hat{\rho}^{A}_{n,m,\pm} &= P^{-1}(\pm) \rho^{A}_{n,m}(0) (c_n\pm s_n) (c_m \pm s_m)\\
\nonumber \\
c_n,s_n &\equiv \cos(M_n\Delta t),\sin(M_n\Delta t)
\end{align}
\end{subequations}
Naturally, this particular choice of the initial state, the interaction Hamiltonian and the projection basis is somewhat arbitrary. The choice can be modified, as long as the qubit precession retains information about $\hat{M}$ and is detected by a projective measurement.

To simulate the CWLM  at a time interval of duration $T$, the step is repeated $N = T/\Delta t$ times. The resulting data set for the measurement results and the density matrices at each step of the time evolution is referred as a quantum trajectory~\cite{Devoret, SiddiqiSingle, SiddiqiEntanglement, WisemanWeakValues}. The averaged quantum evolution is obtained by averaging over the quantum trajectories.\\

If, in addition to the measurement, the system A is subject to Hamiltonian dynamics with Hamiltonian $\hat{H}_A$, this can be included by extending each step with a unitary transformation with the corresponding evolution matrix $\exp(-i \hat{H}_A \Delta t )$.
The error of such separation of the measurement and the Hamiltonian evolution in time scales as $(\Delta t)^2$ and is therefore negligible in the limit of $\Delta t \to 0$.\\

For each run, we obtain a set of $\sigma_i=\pm 1$  measurement outcomes that are almost equally distributed and independent provided that the measurement strength $\hat{M}\Delta t$ of each measurement is small. This is in contrast with an output of a linear detector $V(t)$ that is a continuous number defined for continuous time. It has a white noise spectrum $\langle V(t) V(t')\rangle = \delta(t-t')$. The instant output value has an infinite variance so an actual experimental reading gives the output integrated over a time interval ${\cal T}$
$\bar{V}(t)={\cal T}^{-1}\int_{t}^{t+{\cal T}} d\tau V(\tau)$ that has the finite variance $\langle \bar{V}^2 \rangle = S/{\cal T}$. To simulate the output, we associate
\begin{equation}
\bar{V}(t) = K^{-1} \sum_{i=0}^{K} \sigma_i
\end{equation}
where $K = {\cal T} /\Delta t$, and summation is over $K$ measurement results in the time interval $(t,t+{\cal T})$. The distribution of the sum is normal at $K \gg 1$ so it accurately reproduces the continuous normal-distributed output. Comparing the variances of both sides we conclude that $S=\Delta t$.

We conclude that the qubit can accurately simulate a linear detector provided $\Delta t \ll {\cal T} \ll T$ and $M_n \Delta t  \ll 1$. To provide more accurate estimations, we assume, for the rest of the paper, that the system $A$ is also a qubit and $\hat{M} \equiv M \hat{\Sigma}_z$, $\hat{\Sigma}_i$ being the Pauli matrices in the space of the qubit measured. The eigenvalues of $\hat{M}$ are thus $\pm M$.
We thus provide the linear detection of z-component of the qubit, $\langle V \rangle = 2 M \Delta t \langle \hat{\Sigma}_z\rangle$. For a decisive measurement, the standard deviation of the averaged output signal at the time interval $T$, $\sqrt{S/T}$, should be smaller than the separation $4 M \Delta t$ between the discrete values of the output that correspond to $\langle \hat{\Sigma}_z \rangle = \pm 1$. This gives a typical time scale at which a decisive measurement takes place,
$T_c = (M^2 \Delta t)^{-1}$. Since an interesting simulation would encompass a time interval at least of the order of $T_c$, $N > (M \Delta t)^{-2}$.

A general CWLM is characterized by an inequality \cite{QNoise}
\begin{equation}
S_{out}S_{in} \ge a^2/4
\end{equation}
where $S_{out},S_{in}$ are output and input noises, respectively and $a$ is the linear susceptibility of the detector signal to the input.
Substituting the parameters of our setup, $S_{out}=\Delta t$, $a=2M\Delta t$, $S_M = M^2 \Delta t$, we conclude that our setup simulates an {\it ideal} detector. More general CWLM with non-ideal detector can be simulated if we just add an extra white noise signal to the output, this would lead to an expected deterioration of the measurement quality.

\section{The simple measurement setup: simulation results}
\label{results}
Here, we present the simulation results for a very simple and generic setup. We measure $z$-projection of a qubit pseudospin setting it initially to an equal-weight superposition $|x\rangle$. Owing to the symmetry of the initial condition, $\langle \hat{\Sigma}_z \rangle =0$ at any time. However, at sufficiently long time the superposition is destroyed and the qubit is in one of the states $|+\rangle, |-\rangle$ , this is reflected in the measurement output at sufficiently big durations. One can say that a spontaneous symmetry breaking takes place upon the measurement.
Owing to simplicity of the setup, there are simple and known analytical solutions. The average density matrix satisfies an evolution equation
\begin{equation}
\frac{\partial\hat{\rho}}{\partial t} = \left(\hat{\rho} - \hat{\Sigma}_z \hat{\rho} \hat{\Sigma}_z\right),
\end{equation}
where we measure time in units of $T_c$. The solution that satisfies the initial condition reads $\hat{\rho}(t)= (1+e^{-2 t} \hat{\Sigma}_x)/2$.
One can also evaluate the joint density matrix of the qubit and the measurement outputs. It is convenient for us to use the counting field method \cite{NazWei,  FranquetNaz}. In this method, one considers a time interval $(\tau, \tau +T)$ and solves an evolution equation for the augmented density matrix $\hat{\rho}(\chi)$ in this interval,
\begin{equation}
\label{eq:withcountingfield}
\frac{\partial\hat{\rho}}{\partial t} = -\frac{\chi^2}{8} \hat{\rho} + i\frac{\chi}{2} (\hat{\Sigma}_z \hat{\rho} +\hat{\rho}\hat{\Sigma}_z) - \left(\hat{\rho} - \hat{\Sigma}_z \hat{\rho} \hat{\Sigma}_z\right).
\end{equation}
We normalize the average output in such a way that $v =\pm 1$ for two projections of $\hat{\Sigma}_z$. The joint density matrix $\hat{\rho}(v)$  is then given by
\begin{equation}
\hat{\rho}(v) = \frac{T}{2\pi}\int d\chi e^{-i\chi v T}
\end{equation}
Its trace gives the distribution of the normalized averaged output in this interval.
\subsection{Quantum trajectories}
\begin{figure}[!ht]
\centering
\begin{subfigure}[t]{0.48\textwidth}
\centering
\includegraphics[width=\textwidth]{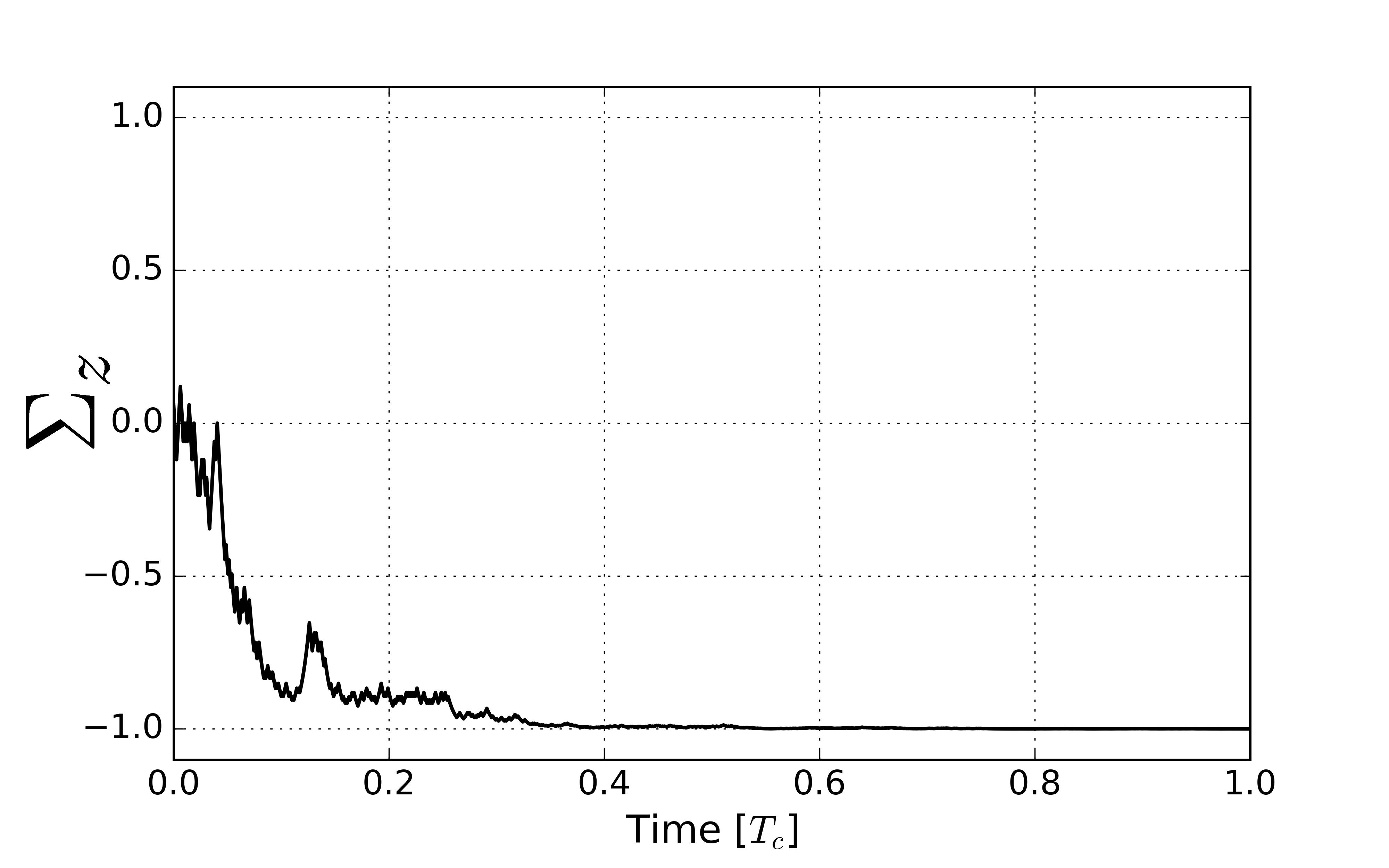}
\caption{Single trajectory.}
\label{fig1a}
\end{subfigure}
~
\begin{subfigure}[t]{0.48\textwidth}
\centering
\includegraphics[width=\textwidth]{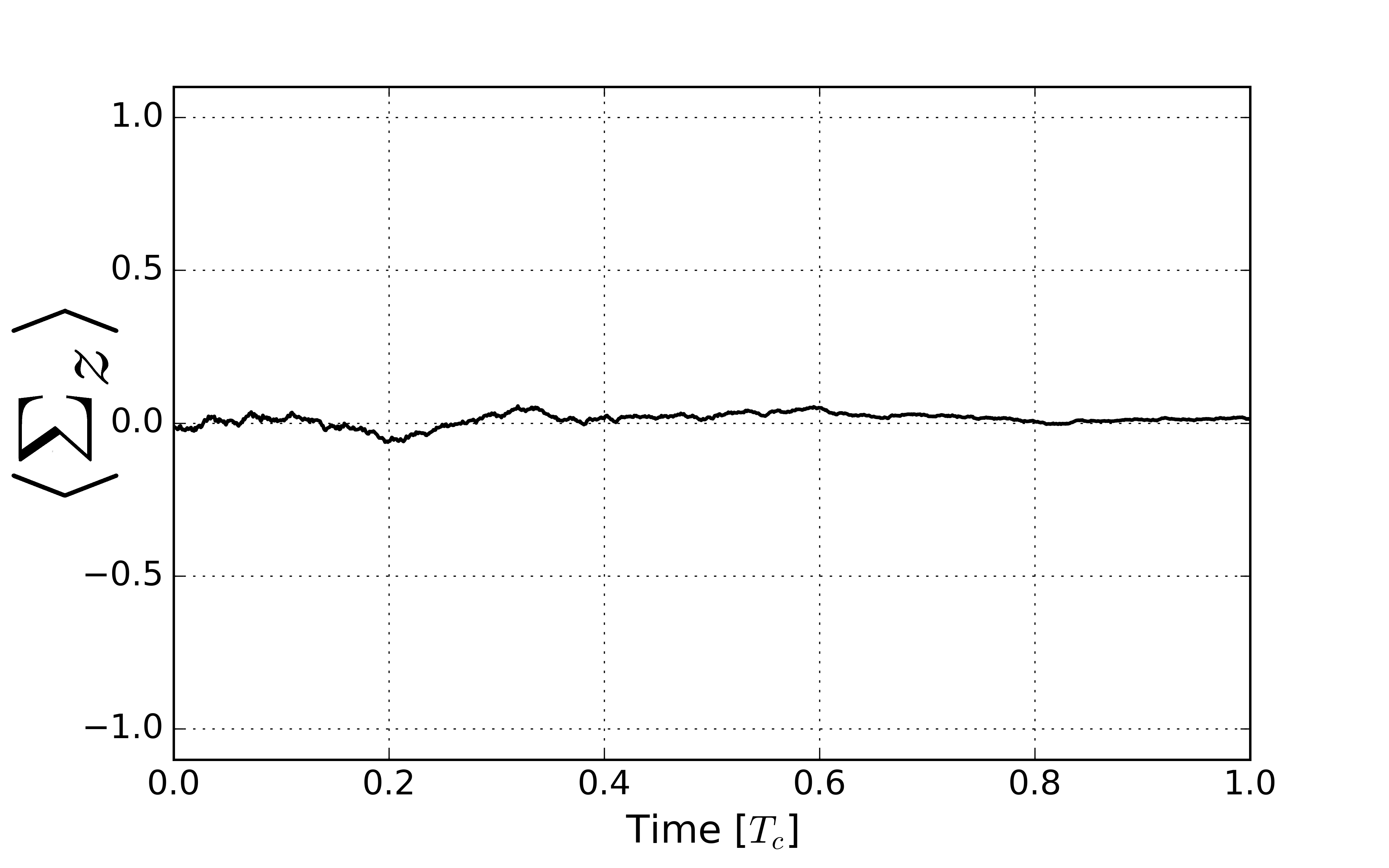}
\caption{Average of 100 trajectories.}
\label{fig1b}
\end{subfigure}
\caption{Quantum trajectories of the  qubit ($\Sigma_z$ is shown )obtained from the simulation. A single trajectory (Fig. (a)) is rather noisy exhibiting sharp jumps induced by the random measurement at each step. A single trajectory gives an information on the {\it random} detector outputs. The averaging over 100  trajectories (Fig. (b)) reproduces the result $\Sigma_z =0$ for the density matrix computed when disregarding the detector outputs. }\label{fig1}
\end{figure}

In Fig.~\ref{fig1} we present a typical output of a  simulation run. The simulations are performed setting the measurement strength to $M\Delta t=0.03$ and choosing the characteristic time scale $T_c=1$. No Hamiltonian evolution is included. Figure~\ref{fig1a} shows a single quantum trajectory of the qubit measured during the CWLM. In order to distinguish various averages, we denote $\Sigma_z(t) = {\rm Tr}(\hat{\Sigma}_z \rho(t))$ the pseudospin component averaged with the density matrix along a single trajectory, while $\langle \Sigma_z(t)\rangle$ denotes the average over the trajectories at a given moment of time.

 We thus plot the $\Sigma_z(t)$. As we see, the projection fluctuates rather wildly, yet approaches $\pm 1$ upon increasing time, so that at sufficiently long time the qubit is projected into either the $\ket{+}$ or the $\ket{-}$ state.\\
For the plot in Fig.~\ref{fig1b}, we run the simulation  100 times and average over all the quantum trajectories. As expected, the contributions of the trajectories with opposite final states compensate each other and  $\langle \Sigma_z\rangle $ approaches zero with $\simeq 10 \%$ deviations.

\subsection{Simulation of the detector signal}
Let us now investigate the detector signal. As described, in our simulation procedure it is obtained by summing up the random results of the projective measurements accumulated during a sampling interval ${\cal T}$. This gives a certain number of detector readings. There is an obvious trade-off between the number of readings and the noise in each reading.

\begin{figure}[!ht]
\centering
\begin{subfigure}[t]{0.48\textwidth}
\centering
\includegraphics[width=\textwidth]{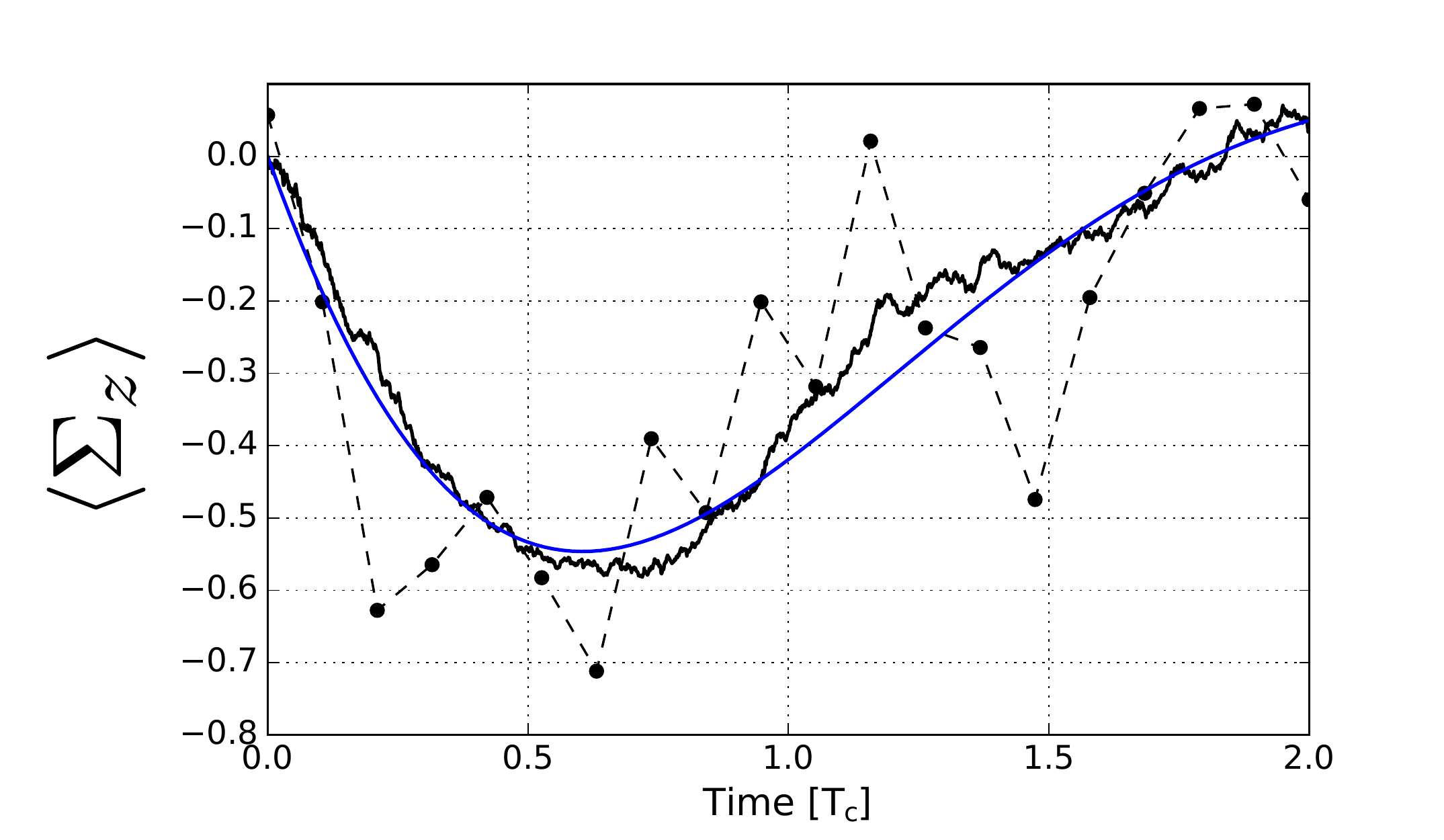}
\caption{Average of 100 trajectories with the sampling interval of ${\cal T}=0.1$ for the detector signal.}
\label{fig2a}
\end{subfigure}
~
\begin{subfigure}[t]{0.48\textwidth}
\centering
\includegraphics[width=\textwidth]{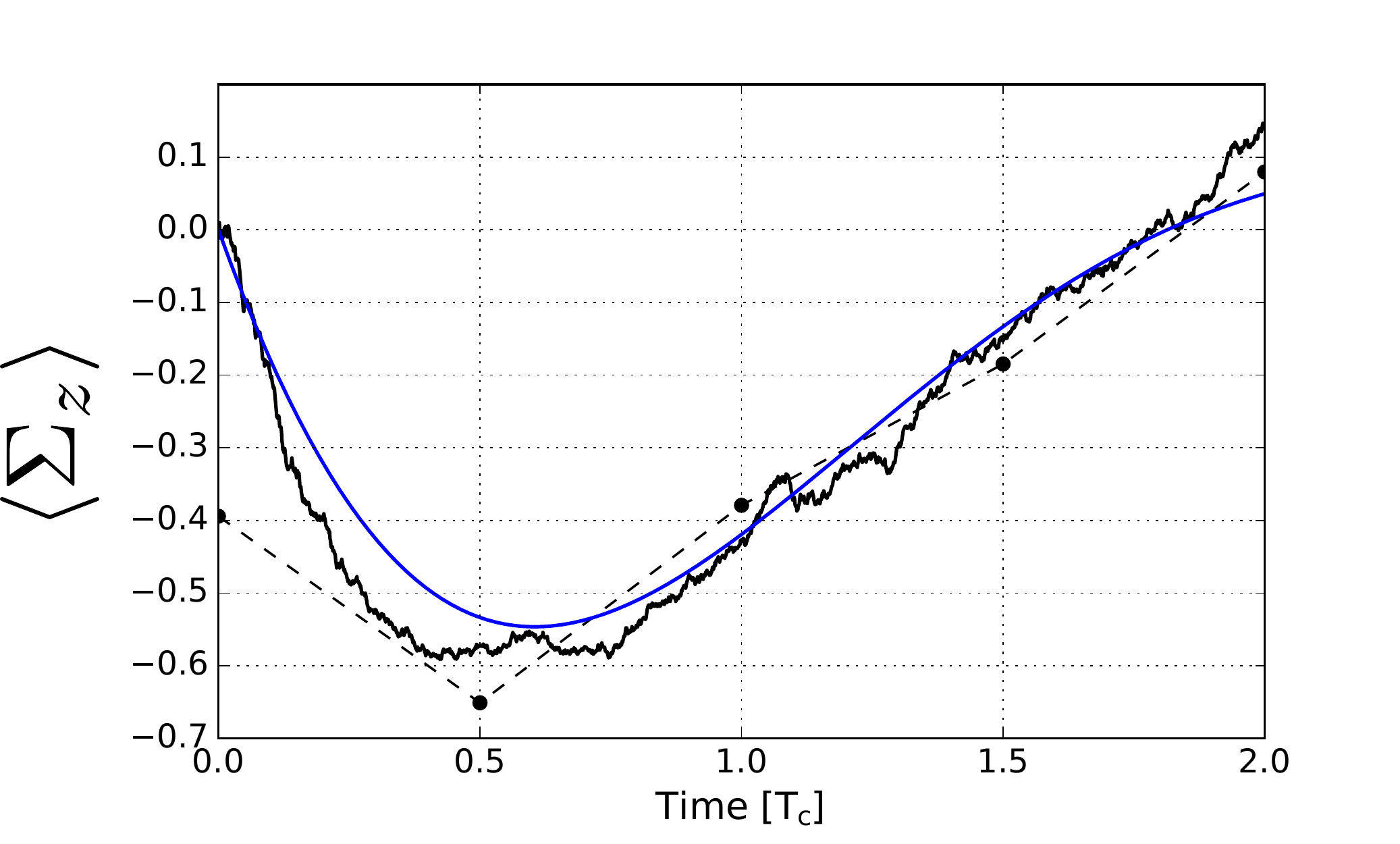}
\caption{Average of 100 trajectories with the sampling interval of ${\cal T}=0.4$ for the detector signal.}
\label{fig2b}
\end{subfigure}
\caption{Averaged qubit trajectories and the corresponding detector signal for different sampling intervals ${\cal T}$. To make the the average detector signal non-zero, we have augmented the dynamics by adding a Hamiltonian $\hat{H}_A=\hbar T_c^{-1}\hat{\sigma}_y$. The duration of the sampling interval ${\cal T}$ controls the noise of the detector readings. The plots illustrate how the average over the trajectories and the detector readings approach $\langle \Sigma_z\rangle(t)$ given by the solid line.}\label{fig2}
\end{figure}
A Hamiltonian $\hat{H}_A=\hbar T_c^{-1}\hat{\sigma}_y$ is added such that the average  $\Sigma_z$ is more "interesting": the Hamiltonian leads to precession of the qubit spin in $x-z$ plane.
For this choice, $\langle \Sigma_z \rangle = 2*\sin(\sqrt{3} t)/\sqrt{3}$. In Fig.~\ref{fig2} we present the average of 100 trajectories and the detector readings for two values of the sampling interval: ${\cal T} =0.1$ (Fig.~\ref{fig2a}) and ${\cal T}=0.4$(Fig.~\ref{fig2b}). We observe that trajectory average is reasonably close to the analytical prediction $\langle \Sigma_z\rangle(t)$. The same holds for the detector readings. However, the correspondence is worse given the same statistics accumulated. This is related to the trade-off mentioned: the readings at short sampling intervals are too noisy, making the interval larger decreases the number of independent detector readings.
\\

\subsection{Results for conditioned output}

\begin{figure}[!ht]
\centering
\begin{subfigure}[t]{0.48\textwidth}
\centering
\includegraphics[width=\textwidth]{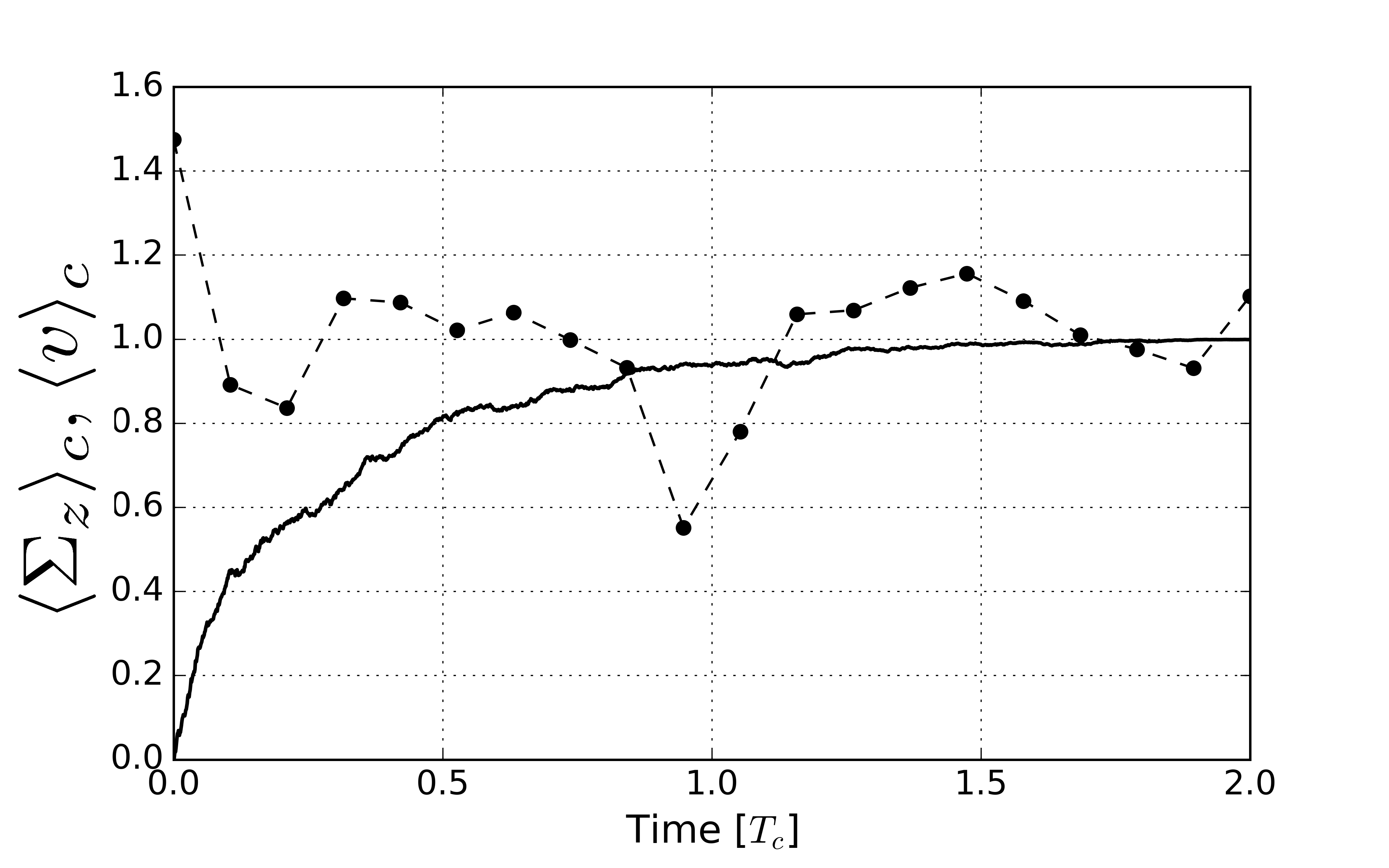}
\caption{The conditional average over 100 trajectories. The detector signal is computed with the sampling interval ${\cal T}=0.1$.}
\label{fig3a}
\end{subfigure}
~
\begin{subfigure}[t]{0.48\textwidth}
\centering
\includegraphics[width=\textwidth]{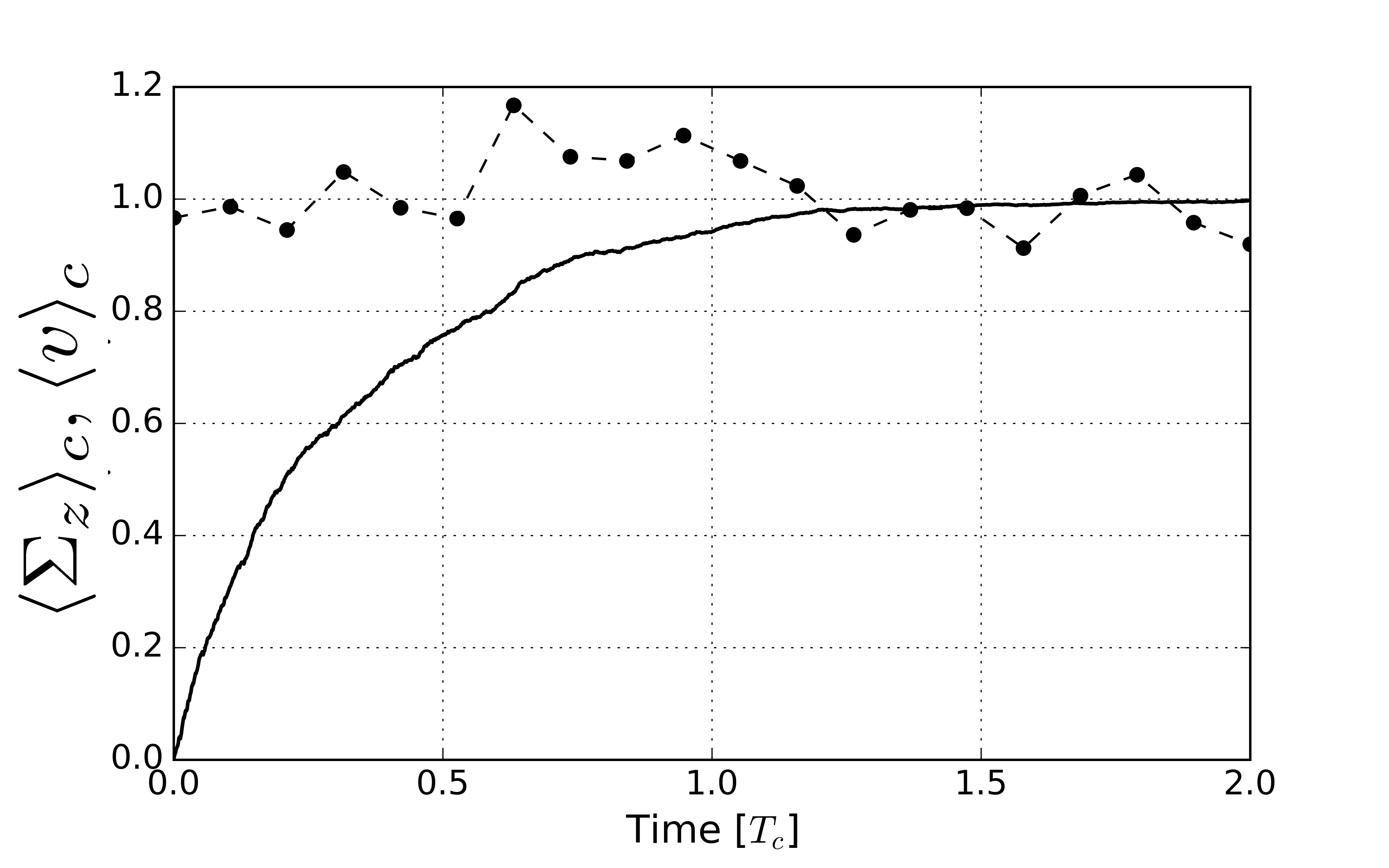}
\caption{The conditional average over 500 trajectories. The detector signal is computed with the sampling interval ${\cal T}=0.1$.}
\label{fig3b}
\end{subfigure}
~\\
\begin{subfigure}[t]{0.48\textwidth}
\centering
\includegraphics[width=\textwidth]{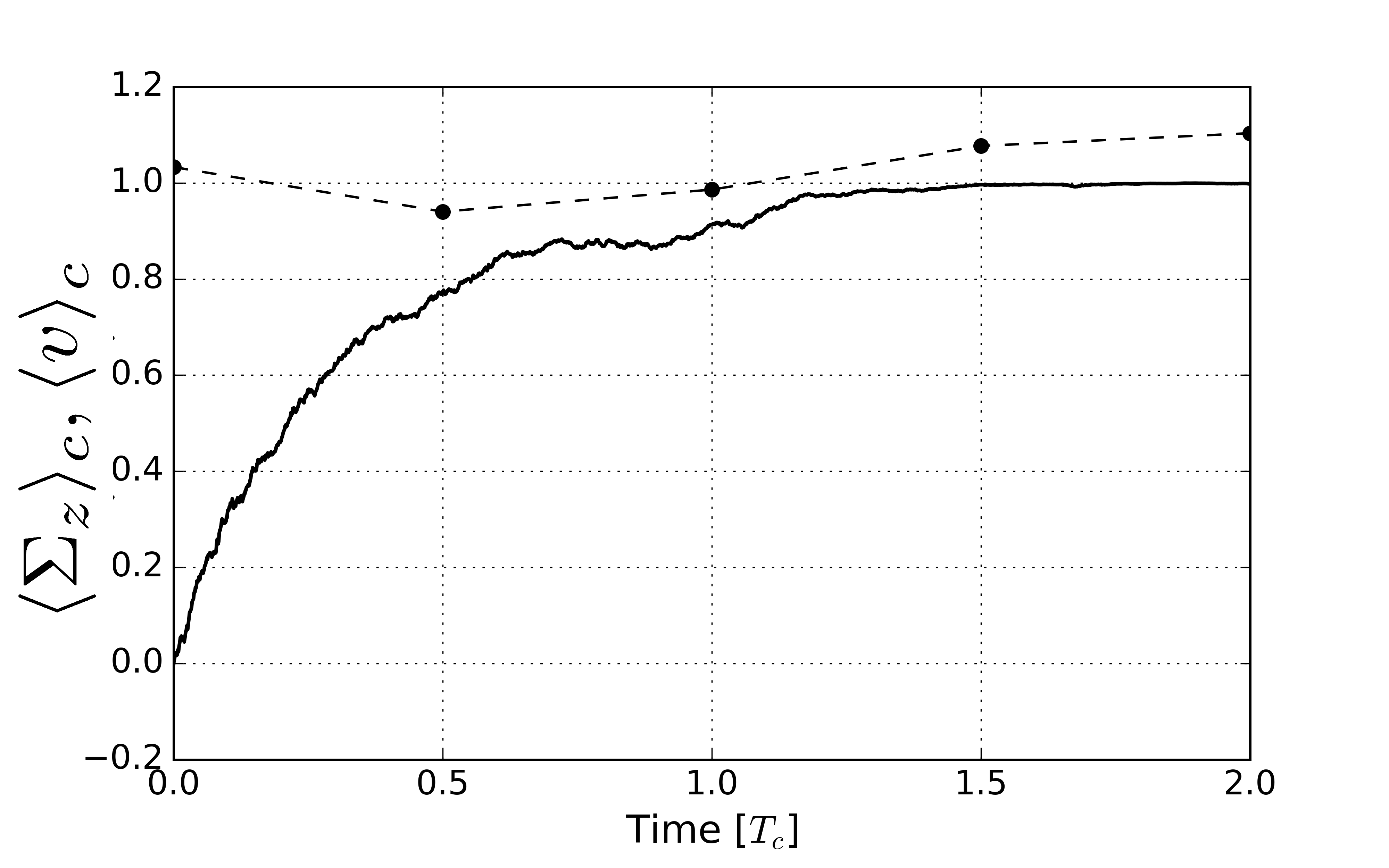}
\caption{The conditional average over 100 trajectories. The detector signal is computed with the sampling interval ${\cal T}=0.4$.}
\label{fig3c}
\end{subfigure}
~
\begin{subfigure}[t]{0.48\textwidth}
\centering
\includegraphics[width=\textwidth]{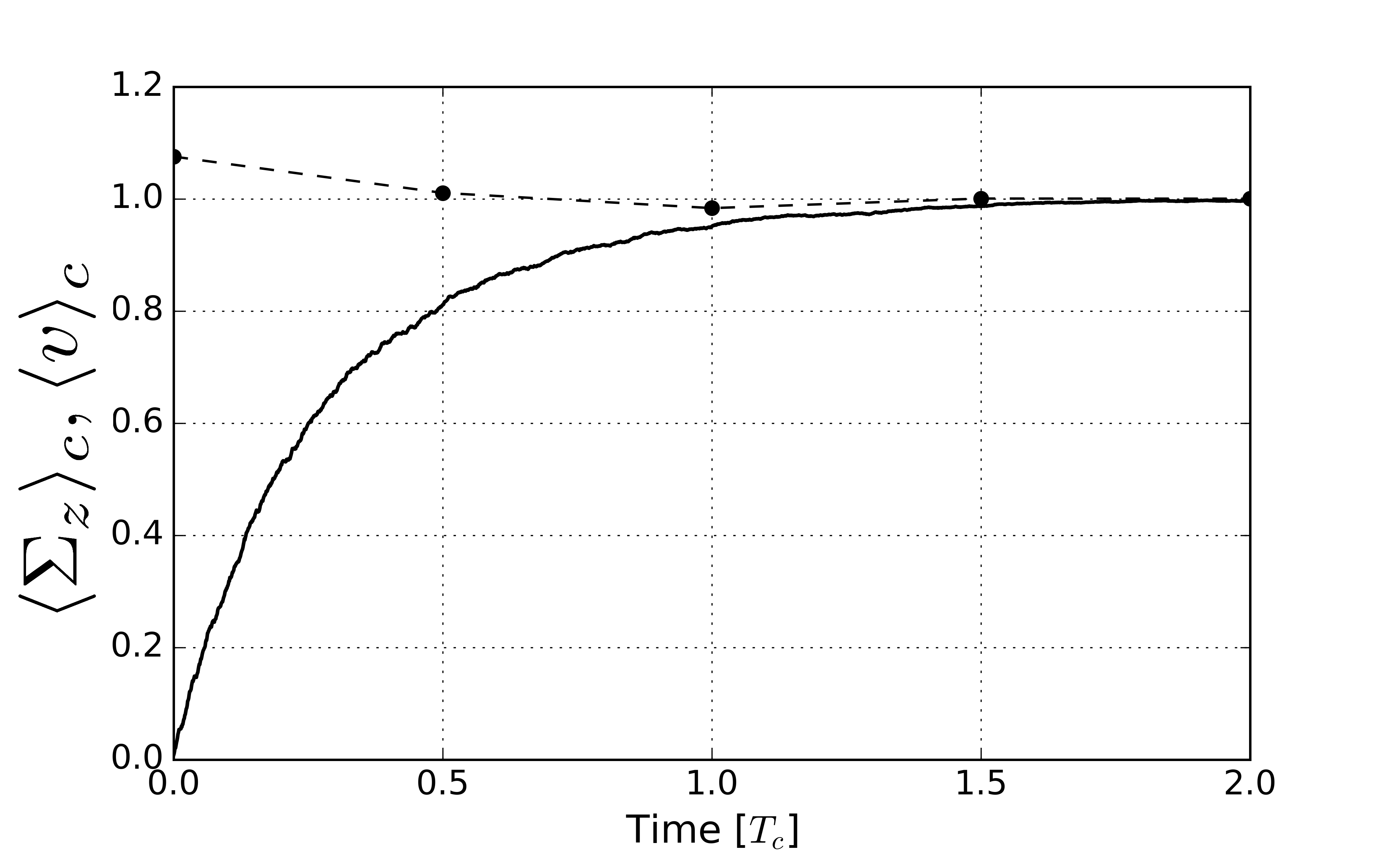}
\caption{The conditional average of 500 trajectories. The detector signal is computed with the sampling interval ${\cal T}=0.4$.}
\label{fig3d}
\end{subfigure}
\caption{The average $\langle{\Sigma}_z(t)\rangle_c $ and the detector signal  $\langle v(t)\rangle_c $conditioned on the final state. The $\langle{\Sigma}_z(t)\rangle_c $ exhibits an expected behavior corresponding to transition from the equal-weight superposition at $t=0$ where  ${\Sigma}_z=0$ to the final $\ket{\pm}$ where $\langle{\Sigma}_z(t)\rangle_c =1$ at a time scale $\simeq T_c$.
Rather counter intuitively, the average output does not follow $\langle{\Sigma}_z(t)\rangle_c$, and, with numerical accuracy, does not depend on time, $\langle v(t)\rangle_c=1$. It looks like the qubit "knows" from very beginning that it is in a final state.}
\label{fig3}
\end{figure}

An interesting behaviour of the detector output can be seen in conditioned measurements \cite{FranquetNaz}. For our setup, it is natural to condition the quantum trajectories on their asymptotic values at long times where the corresponding $\Sigma_z$ sticks to $\pm 1$.This is equivalent to a post-selection to the states $\ket{+}$ or $\ket{+}$. So we accumulate the statistics of the quantum trajectories and the corresponding detector outputs taking the values of $\Sigma_z(t)$ and $v(t)$ with the sign of $\Sigma_z(\infty)$ (or, equivalently, $v(\infty)$, since the output corresponds to the state at $t \to \infty$). We disregard the Hamiltonian dynamics, $\hat{H}_A=0$. \\
In Fig.~\ref{fig3} we present these conditional averages of $\Sigma_z(t)$ and $v(t)$, $\langle \Sigma_z(t) \rangle_c$,$\langle v(t) \rangle_c$\\
In Fig. \ref{fig3a} and \ref{fig3b} the sampling interval is chosen ${\cal T}=0.1$ while in Fig.~\ref{fig3c} and \ref{fig3d} we use ${\cal T}=0.4$. We average over 100 post-selected trajectories in Fig.~\ref{fig3a} and \ref{fig3c} , and over 500 post-selected trajectories in Fig.~\ref{fig3b} and \ref{fig3d}.\\

Let us discuss first the conditioned average of $\Sigma(z)$, $\langle{\Sigma}_z(t)\rangle_c$. As one may expect, it starts at 0 at $t=0$ where qubit is in the equal-weight superposition and approaches $1$ at the time scale $\simeq T_c$. Collecting statistics of 20000 trajectories, we have shown that with $`10^{-2}$ relative accuracy $\langle{\Sigma}_z(t)\rangle_c = {\rm tanh}(f(t))$, $f(t)= t(1.15 +2.8/(1+4.2t))$.

Generally, one may expect that the average detector signal follows $\langle{\Sigma}_z(t)\rangle_c$. We have seen that this is the case for unconditional average. Rather surprisingly, it does not. Moreover, the average signal does not depend on time, $\langle v(t)\rangle_c=1$. It looks like the qubit initially is not in a superposition, but just from the beginning is already in one of $\ket{\pm}$ states, and this state does not change during the measurement. The observed conditioned output would be the same as from a classical bit that is randomly put to one of the two states in the beginning.

It should be possible to confirm such simple result analytically.
Indeed, it follows from a straightforward calculation that employs the formalism introduced in~\cite{NazWei,  FranquetNaz}. Let us collect detector output during two time intervals: first one of duration $t_1$ and the second one that follows the first and has the duration $t_2$. To find the distribution of two outputs $v_{1,2}$, we need to solve Eq. \ref{eq:withcountingfield} for a time-dependent $\chi(t)$ that takes values $\chi_{1,2}$ in the intervals and is zero otherwise.
The  distribution is computed from Fourier transform of ${\rm Tr}[\rho(\chi_{1,2}]$ and reads
\begin{equation}
P(v_1,v_2)=\frac{\sqrt{t_1t_2}}{\pi}\sum_{\pm} e^{-2 (v_1\pm 1)^2 t_1}e^{-2 (v_2\pm 1)^2 t_2}
\end{equation}
It does not depend on the start time moments of the interval but only on their durations.
To adjust this general expression to our situation,
we increase $t_2$ restricting $v_2$ to $\pm 1$. The conditional probability then becomes

\begin{equation}
P(v_1|v_2=1)=\sqrt{\frac{2t_1}{\pi}}e^{-2(v_1-1)^2 t_1}.
\end{equation}
Apparently, $\langle v(t)\rangle_c=1$ does not depend on time, in agreement with the numerical results.

\subsection{Decision time distribution}
\begin{figure}[!ht]
\centering
\includegraphics[width=0.9\textwidth]{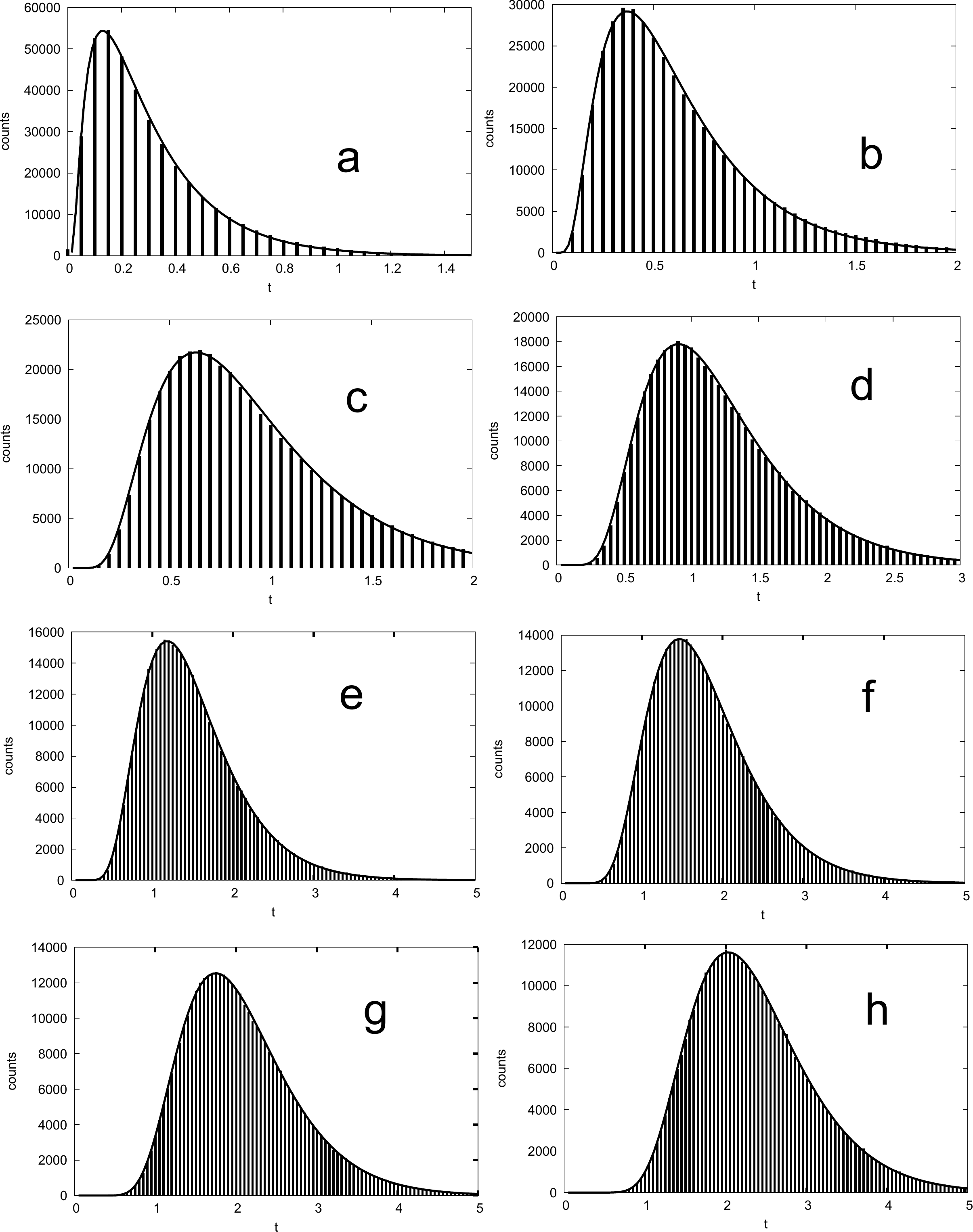}
\caption{Histograms of the decision time at various values of the threshold parameter $h$. (a) $h=0.1$, (b) $h=0.01$, (c) $h=0.001$ , (d) $h=10^{-4}$, (e) $h=10^{-5}$,(f) $h=10^{-6}$,(g) $h=10^{-7}$, (h) $h=10^{-8}$. The body of the distribution shifts to longer times upon decreasing $h$ and the variance decreases slightly. The fit is made with an exponential function of the form $ce^{-a/t-bt}$, $a,b$ being free parameters and $c$ being set by the normalization. The fits are plotted with solid lines. }
\label{fig4}
\end{figure}

Let us consider a knowledgeable observer who has access to all the results of the projective measurements of the detector. With this, and with the known initial condition he is able to reconstruct the density matrix along an individual quantum trajectory and monitor it in time.
Suppose he needs to decide upon the final state of the qubit as soon as possible. He does this by monitoring $\Sigma_z(t)$. Whilst its absolute value reaches a certain threshold $|\Sigma_z| = 1-h$, he makes the decision based on the sign of $\Sigma_z$. We note that the decision may be wrong, and further evolution along the trajectory would bring the qubit to the opposite quantum state. Association $p_{\pm} = (1+\pm \Sigma(z))/2$ suggests that the probability of error is $0.5 h$, and this is confirmed by our numerical simulations. Thus the decision is well-based in the limit of small $h$. So-defined decision time is thus a random quantity, its distribution depending on $h$.
 This distribution is useful for a less knowledgeable and less devoted observer, who just wishes to quantify a time required for the qubit to come to a certain state with sufficiently high probability.
 \\
We have evaluated the distribution numerically at various small threshold values $h$ collecting the statistics of $4\cdot 10^5$ trajectories. We have made the histograms and fitted their shape. The results are presented in Fig.~\ref{fig4}.

$$
t_o = -0.12 \ln h -0.2
b = 2.0 - 6 \ln h $$
As expected, the body of the distribution shifts to longer times upon decreasing $h$, this is accompanied by an increase in the variance. We choose to fit the distribution shape with a rather arbitrary  function $ce^{-a/t-bt}$ which is exponentially small at short and long times, $a,b$ being free coefficients and $c$ being fixed by the normalization. The fits are excellent, especially at smaller $h$. The values of the coefficients $a$, $b$ for different $h$ are given in the table \ref{table1}. \\

\begin{table}
\begin{tabular}{c | c c }
$h$ & $a$ & $b$  \\
\hline
$10^{-1}$ & 0.0820617 & 4.88529 \\
$10^{-2}$ & 0.452684 & 3.28453\\
$10^{-3}$ & 1.13362240458 & 2.84497317856 \\
$10^{-4}$ & 2.14055480776 & 2.61165068723 \\
$10^{-5}$ & 3.5115442592 & 2.49436474245 \\
$10^{-6}$ & 5.17495025523 & 2.41115342218\\
$10^{-7}$ & 7.20739094438 & 2.35187767864 \\
$10^{-8}$ & 9.57228930443 & 2.31735035473 \\
\end{tabular}
\caption{The fit coefficients for the decision time distribution. \label{table1}}
\end{table}

The most probable decision time $t_p \equiv \sqrt{a/b}$ that corresponds to the maximum of the distribution can be fitted well with $t_p = - \ln(2.3 h)$. We note that this is rather short time in comparison with the life-time of the superposition. Since $\langle \Sigma_x(t)\rangle = e^{-2t}$, $\langle \Sigma_x (t_p) \rangle \approx (2.3 h)^{1/4} \gg h$, although a naive expectation would be  $\langle \Sigma_x (t_p)\rangle  \simeq h$. The distribution has a prominent exponential tail at $t \to \infty$. The corresponding coefficient $b$ can be neatly fitted with $ b = 2.0 -6/\ln h$ thus approaching $2$ at small thresholds. This is probably the manifestation of the superposition life-time.
We note that although the variance $\approx 0.25 \sqrt{a/b^3} \approx t_p/8$ grows with decreasing $h$, the relative variance $\approx (8 t_p)^{-1}$ actually decreases resulting in a concentrated distribution.
This permits an accurate quantification of the expected decision time for small $h$ and corresponding error probability $h/2$.

\section{Results on the feedback scheme}
\label{sec:feedback}

One can think of further technological developments whereby the information collected in course of the CWLM is used to manipulate the measured system. Modern qubit implementations make it realistic. Here we consider a simple example of such feedback scheme.

The measurement destroys the superposition and brings the qubit to a certain final state. The detector shows what state is reached. Let us note that the state can be "corrected": the qubit can be brought back to the initial superposition by a unitary manipulation, rotation about $Y$ axis.
The rotation angle, however, does depend on the state reached. General rotation by angle $\alpha$ is given by a unitary matrix $\hat{U}(\alpha) = \cos\alpha + i \Sigma_y \sin\alpha$. We see that $\ket{\pm}$ state is corrected by $\hat{U}(\pm \pi/4)$.

Let us devise a simple feedback scheme with a goal to keep the qubit in the equal-weight superposition while being measured. It works as follows. We collect the detector output during a time interval $T_f$. We use the reading $v$ to decide which rotation we apply.
The simplest decision scheme utilizes a reaction threshold $I$:
no correcting manipulation takes place if $|v|<I$, otherwise the rotation $\hat{U}({\rm sgn}(I) \pi/4)$ is applied. Alternatively, the rotation angle is
\begin{equation}
\label{eq:correctingangle}
\alpha(v) = {\rm sgn}(v) \Theta(|v|>I).
\end{equation}
 Then the feedback cycle is repeated again and again: the collection of the output at a time interval $T_f$ is followed by a correcting rotation.

If the collection time $T_f \gg T_c$, the correction to the superposition will be exact if $I<1$. However, the superposition will be destroyed at the time scale of $T_c$ and will persist for only a small fraction of the cycle. In the opposite limit $T_f \ll T_c$ the superposition will not be destroyed during the cycle. However, the output collected at such small time interval will exhibit large fluctuations and will hardly reflect the state measured. This will make the correction very inefficient. As a criterion for a good feedback, we take the average value of $\Sigma_x(t)$ over the whole cycle, $\bar{\Sigma_x} \equiv T_f^{-1}\int_0^{T_f}dt\langle\Sigma_x (t)\rangle$. This value will depend on
$T_f$ and $I$, and we will find the optimal values of these parameters.

\subsection{Analytics}
\label{sec:analyticalfeedback}
Owing to the simplicity of the scheme, we can find analytical expressions for $\bar{\Sigma}_x$.
We note that the solution for the density matrix must be periodic in time with the period $T_f$. Let $\hat{\rho}_a$ be the density matrix of the qubit right after the correction. It evolves on the time interval $T_f$ to the joint density matrix $\hat{\rho}(v)$ of the qubit and output. This can be found by solving Eq. \ref{eq:withcountingfield} with the initial condition ${\hat\rho}_a$.
Applying the output-dependent correction to the joint density matrix, we return to $\hat{\rho}_a$,
\begin{equation}
\hat{\rho_a} = \int dv \hat{U}(\alpha(v)) \hat{\rho}(v)\hat{U}(\alpha(v))^{-1}(\alpha(v))
\end{equation}
This forms a closed self-consistency equation for $\hat{\rho_a}$ to solve.
In our situation, owing to symmetry, we can seek for $\hat{\rho}_a$ in the form $\hat{\rho}_a = (1+\rho_x \hat{\Sigma}_x)/2$. The joint density matrix takes the form
\begin{equation}
\hat{\rho}(v) = \sum_{\pm}G_{\pm}\frac{1+\hat{\Sigma_z}}{4} +
e^{-2T_f}G \rho_x \Sigma_x,
\end{equation}
where $G(v)\equiv (2T_f/\pi)^{1/2} \exp(-2 T_f v^2)$,
$G_{\pm}(v) = G(v \mp 1)$.

The self-consistency equation reads
\begin{eqnarray}
\rho_x &=& A+B \rho_x \\
A&\equiv& \int dv \frac{G_+-G_-}{2} \sin(2\alpha(v)) \\
&=& \frac{1}{2}(erf((I+1)\sqrt{2T_f})- erf((I-1)\sqrt{2T_f})) \\
B&\equiv& \int dv e^{-2T_f}\int G \cos (2\alpha(v)) = e^{-2T_f}{\rm erf}(I\sqrt{2T_f})
\end{eqnarray}
The time-averaged $x$-projection is computed as
\begin{equation}
\bar{\Sigma}_x = \rho_x \frac{1-e^{2T_f}}{2T_f} = \frac{A}{1-B} \frac{1-e^{2T_f}}{2T_f}
\end{equation}

\begin{figure}[!ht]
\centering
\includegraphics[width=0.5\textwidth]{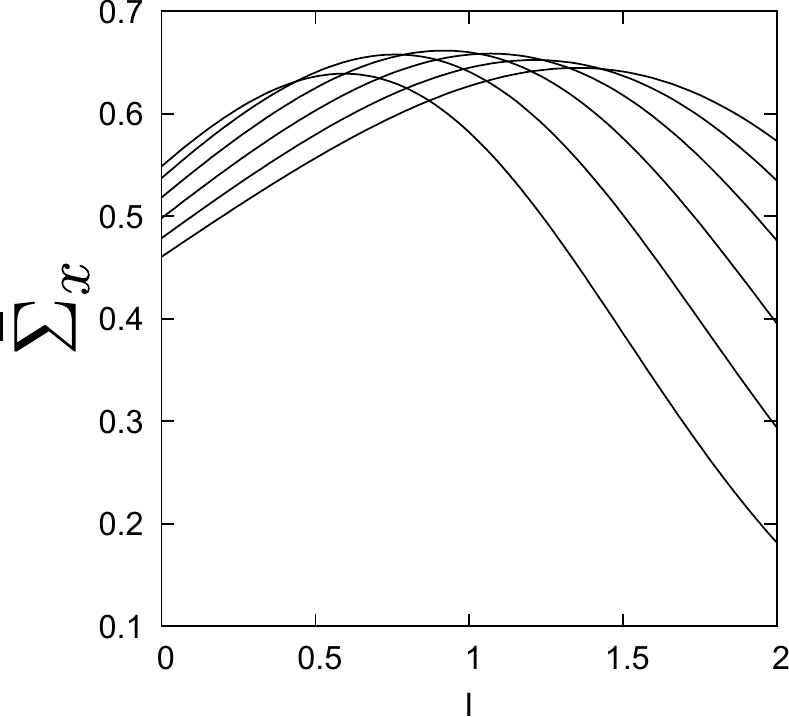}
\caption{The efficiency $\bar{\Sigma}_x$ versus the reaction threshold $I$ for a set of collection times $T_f$. The $T_f$  takes the values $1/3,1/4, 1/5, 1/6, 1/7, 1/8$ from the lower to the upper curve at $I=2$, respectively. The curves come in opposite order at $I=0$. The plot shows that the efficiency close to $2/3$ can be achieved in a wide region of $I$ and $T_f$.}
\label{fig:feedbackanalytics}
\end{figure}

In Fig. \ref{fig:feedbackanalytics} we plot $\bar{\Sigma}_x$ versus $I$ for a set of $T_f$. The curves reach maximum at some intermediate value of $I$. More detailed optimization shows that the maximum value of $\bar{\Sigma}_x =0.661$ is achieved at $I=0.88$, $T_f =0.21$. The average value of spin immediately after the correction is higher, $\langle \Sigma_x \rangle_a = 0.81$ for the optimal settings.  We see from the plot that close values of $\bar{\Sigma}_z$ are achieved in a rather wide window of $I$ and $T_f$. This is a rather large value given the primitive feedback scheme in use. More elaborated feedback schemes may improve this even further.

\subsection{Numerical results
} \label{sec:numfeedback}
We investigate the feedback scheme numerically with the tool described. The simulation proceeds by time intervals of duration $T_f$. The detector output is collected during the interval, and the correcting rotation about $y$ axis is applied depending on the resulting output in accordance with Eq. \ref{eq:correctingangle}. We always start with the equal-weight superposition at $t=0$ and collect the quantum trajectories along with the detector readings. Some time is required for the simulation to achieve a steady state where the averages are periodic. We have found that in the range of $T_f$ explored this time is of the order of 5-7 cycles irrespective of the cycle duration.

We explore and illustrate numerically the effect of the reaction threshold $I$ and the collection time $T_f$ on the performance of the feedback scheme and find numerically the optimal settings $I$, $T_f$ that maximize this performance.\\

\begin{figure}[!ht]
\centering
\includegraphics[scale=0.4]{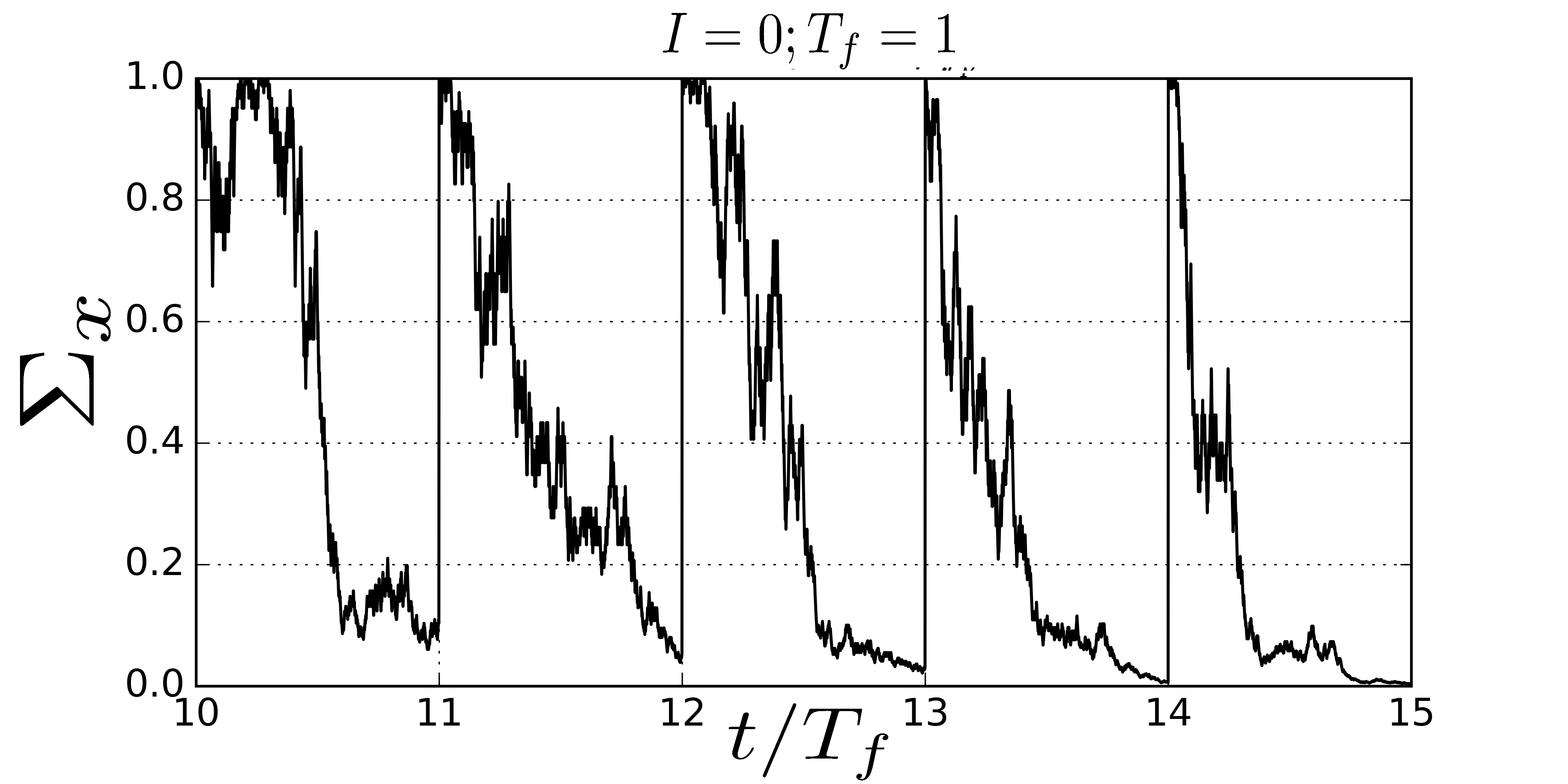}
\caption{A single quantum trajectory ${\Sigma}_x(t)$ of the qubit subject to the feedback. The collection time is set to a rather large value $T_f=1$, $I=0$. At these settings, the superposition is strongly suppressed within the collection time. This provides an accurate measurement and efficient correction to the target superposition state. }
\label{fig5}
\end{figure}

Figure~\ref{fig5} gives an example of a single quantum trajectory. We plot $\ex{\hat{\sigma}_x}$ versus time for 5 collection interval. The collection time is set to a rather large value $T_f=1$. The superposition is essentially suppressed during this time so the measurement of the final state is accurate and the resulting correction is accurate. We see the $\ex{\hat{\sigma}_x}$ coming back to $\approx 1$ any time after the correction. We also see strong and fast fluctuations of $\ex{\hat{\sigma}_x}$ in time.

\begin{figure}[!ht]
\centering
\begin{subfigure}[t]{0.48\textwidth}
\centering
\includegraphics[width=\textwidth]{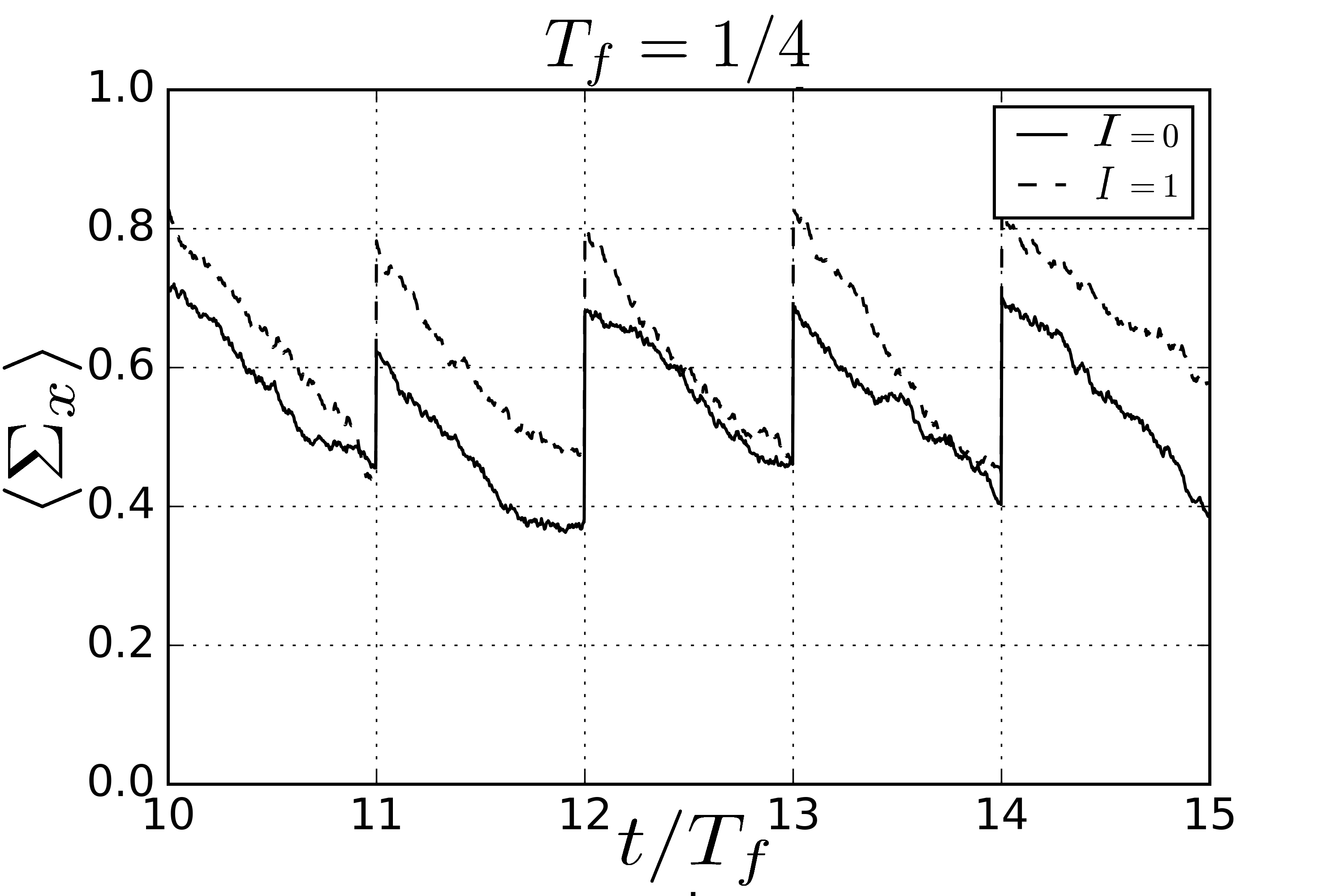}
\caption{The average of $\ex{{\Sigma}_x(t)}$ over 50 trajectories for $T_f=0.25$ and two values of the reaction threshold, $I=0$ (solid) and $I=1$ (dashed).
For the latter setting, the correction is not applied for small values of the collected output $v$. Apparently, this improves the feedback performance. It is better to do nothing then wrong.
}
\label{fig6a}
\end{subfigure}
~
\begin{subfigure}[t]{0.48\textwidth}
\centering
\includegraphics[width=\textwidth]{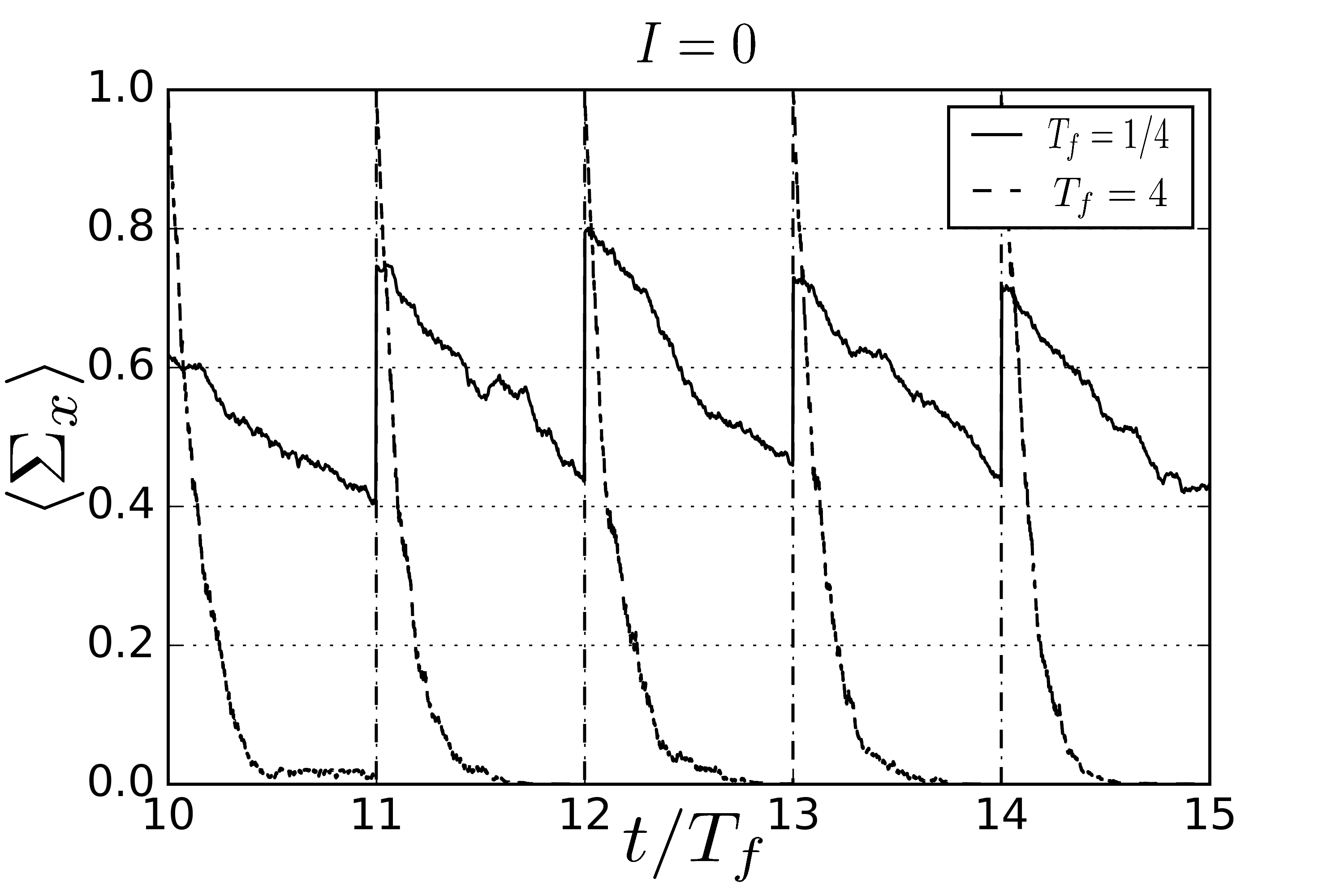}
\caption{The average $\ex{\hat{\Sigma}_x(t)}$ over 50 trajectories for $I=0$ and two values of $T_f$, $T_f =4$ (dashed) and $T_f =1/4$ (solid). Long collection time improves the efficiency of the correction but does not keep the superposition. $T_f=0.25$ is close to optimal settings providing a good trade-off between the superposition decay and the variance of the output collected. }
\label{fig6b}
\end{subfigure}
\caption{The effects of the parameters $I$ and $T_f$ on the feedback performance illustrated with numerical simulations.}\label{fig6}
\end{figure}

To suppress these fluctuations, we plot in the subsequent figures the averages over 50 quantum trajectories.
To illustrate the effect of the parameters $I$ and $T_f$ on the dynamics of the qubit under feedback, we present in Fig.~\ref{fig6} such averages for different parameter values. In Figure~\ref{fig6a} we set $T_f=1/4$ and compare the results for $I=0$ (solid) and $I=1$ (dashed). One can see the improved performance in the latter case: the qubit is closer to the target equal-weight superposition. At this choice of $I$, no correction is applied if the collected output $|v|<1$ and its reduced value does not indicate a certain $z$-projection. At these settings, this happens in approximately $1/3$ of the cases. Apparently, the rule "it is better to do nothing than to do wrong" works here well. The reduced values of the output signal a relatively high value of $\Sigma_x$ that does not have to be corrected. \\
 In Fig. ~\ref{fig6b} we set $I=0$ and plot the  average $\ex{{\Sigma}_x(t)}$  for two collection times: $T_f=1/4$ (solid curve) and $T_f=4$ (dashed). For the long collection time, we observe almost complete decay of the superposition and accurate correction to the target superposition at each feedback cycle. For the shorter correction time, the correction at each cycle is by far less accurate, but the superposition does not decay much and is big in average.\\

\begin{figure}[!ht]
\centering
\includegraphics[scale=0.6]{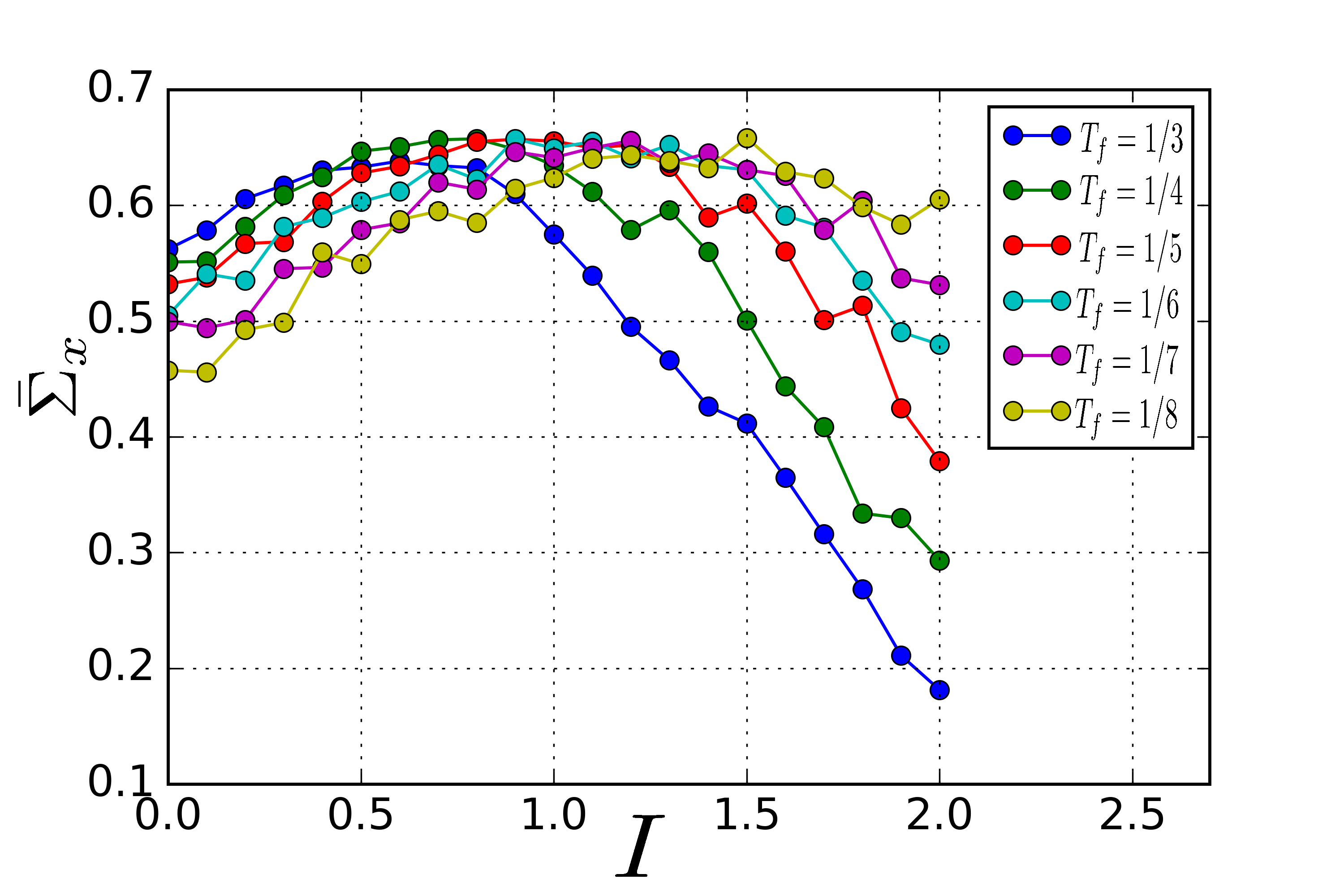}
\caption{(Color online) The feedback efficiency $\bar{\Sigma}_x$ in a wide region of the reaction thresholds $I$ and collection times $T_f$. Each data point is computed by averaging of 500 quantum trajectories over time interval of 100 $T_f$. An efficiency $\simeq 0.6$ is observed in a wide parameter range.}
\label{fig7}
\end{figure}

We explore  the feedback efficiency $\bar{\Sigma}_x$ in a wide range of $I$, $T_f$. We present the results in Fig. ~\ref{fig7} where each point corresponds to averaging over  500 quantum trajectories during 100 collection time intervals. Apart from the remaining noise, these numerical data are in agreement with the results of the analytical calculation presented in Fig.\ref{fig:feedbackanalytics}. Each data set at fixed $T_f$ exhibit a maximum in efficiency at some intermediate value of $I$.
The exact optimization settings are difficult to see since the similar efficiency close to $2/3$ is reached in a wide region of the parameters. We employ the numerical iterative optimization procedure that bring us to the values $\bar{\Sigma}_x =0.66$  at $I=0.9$, $T_f =0.2$ that is close to the settings obtained from numerical analysis. This proves that the simulation tool in use can be efficiently implemented for the analysis of more sophisticated and efficient feedback schemes that are too complicated to be treated analytically.

The efficiency of the feedback scheme can be definitely improved, for instance, by choosing the rotation angle $\alpha(v)$ in a more flexible way and taking into account the detector outcomes from the previous collection intervals.\\

\section{Conclusion}
\label{conclusion}

In this paper, we explore the peculiarities of a continuous weak linear measurement in a simple but generic setup. We develop an efficient numerical simulation tool that generates single quantum trajectories along with the corresponding detector signal. We study the statistics of the trajectories and outputs with and without post-selection.

We prove numerically that the detector output conditioned on the final state does not depend on time and does not follow the average $\Sigma_z(t)$. Seemingly this implies that the measured qubit "knows" from very beginning of the measurement in which final state it is. To investigate this further, we study the statistics of the decision times. We have found an accurate fit for corresponding distribution and revealed that the decision time is commonly much shorter than the life-time of the superposition.
  present a procedural way of simulating and modelling a continuous measurement of an arbitrary quantum system. In particular a CWLM of a qubit is modelled using another qubit as a linear detector.\\
This simple description gives insight into the interplay of the measured system and the detector system from a quantum point of view. It describes the translation of discrete quantum information to a continuous classical signal.\\
While similar methods have been successfully used to described such measurement scenarios~\cite{Devoret, SiddiqiSingle, SiddiqiEntanglement, WisemanWeakValues}, our method allows not only to describe the measured system dynamics but also the detector system signal. How this signal is constructed and in which circumstances can correspond to a real integrated signal of an experimental run.\\
We have also presented and investigated a simple feedback scheme where the measurement results are used to keep the qubit in the initial superposition state despite being measured. Despite the simplicity, the feedback scheme can be tuned to provide rather high efficiency ${\bar \Sigma}_x =0.66$. The feedback can be further improved and sophisticated.

The results obtained are relevant in the context of experimental situations where continuous weak linear measurement is used and for the design of interesting quantum feedback schemes and measuring protocols.

\begin{acknowledgments}
This work was supported by the Netherlands Organization for Scientific Research (NWO/OCW), as part of the Frontiers of Nanoscience program.
\end{acknowledgments}

\bibliography{paper_clean.bib}

\providecommand{\noopsort}[1]{}\providecommand{\singleletter}[1]{#1}%
\begin{thebibliography}{24}%
\makeatletter
\providecommand \@ifxundefined [1]{%
 \@ifx{#1\undefined}
}%
\providecommand \@ifnum [1]{%
 \ifnum #1\expandafter \@firstoftwo
 \else \expandafter \@secondoftwo
 \fi
}%
\providecommand \@ifx [1]{%
 \ifx #1\expandafter \@firstoftwo
 \else \expandafter \@secondoftwo
 \fi
}%
\providecommand \natexlab [1]{#1}%
\providecommand \enquote  [1]{``#1''}%
\providecommand \bibnamefont  [1]{#1}%
\providecommand \bibfnamefont [1]{#1}%
\providecommand \citenamefont [1]{#1}%
\providecommand \href@noop [0]{\@secondoftwo}%
\providecommand \href [0]{\begingroup \@sanitize@url \@href}%
\providecommand \@href[1]{\@@startlink{#1}\@@href}%
\providecommand \@@href[1]{\endgroup#1\@@endlink}%
\providecommand \@sanitize@url [0]{\catcode `\\12\catcode `\$12\catcode
  `\&12\catcode `\#12\catcode `\^12\catcode `\_12\catcode `\%12\relax}%
\providecommand \@@startlink[1]{}%
\providecommand \@@endlink[0]{}%
\providecommand \url  [0]{\begingroup\@sanitize@url \@url }%
\providecommand \@url [1]{\endgroup\@href {#1}{\urlprefix }}%
\providecommand \urlprefix  [0]{URL }%
\providecommand \Eprint [0]{\href }%
\providecommand \doibase [0]{http://dx.doi.org/}%
\providecommand \selectlanguage [0]{\@gobble}%
\providecommand \bibinfo  [0]{\@secondoftwo}%
\providecommand \bibfield  [0]{\@secondoftwo}%
\providecommand \translation [1]{[#1]}%
\providecommand \BibitemOpen [0]{}%
\providecommand \bibitemStop [0]{}%
\providecommand \bibitemNoStop [0]{.\EOS\space}%
\providecommand \EOS [0]{\spacefactor3000\relax}%
\providecommand \BibitemShut  [1]{\csname bibitem#1\endcsname}%
\let\auto@bib@innerbib\@empty
\bibitem [{\citenamefont {Nielsen}\ and\ \citenamefont
  {Chuang}(2000)}]{NielsenChuang}%
  \BibitemOpen
  \bibfield  {author} {\bibinfo {author} {\bibfnamefont {M.}~\bibnamefont
  {Nielsen}}\ and\ \bibinfo {author} {\bibfnamefont {I.}~\bibnamefont
  {Chuang}},\ }\href@noop {} {\emph {\bibinfo {title} {Quantum Computation and
  Quantum Information}}}\ (\bibinfo  {publisher} {Cambridge University Press},\
  \bibinfo {year} {2000})\BibitemShut {NoStop}%
\bibitem [{\citenamefont {Mensky}(1991)}]{CWLM0}%
  \BibitemOpen
  \bibfield  {author} {\bibinfo {author} {\bibfnamefont {M.~B.}\ \bibnamefont
  {Mensky}},\ }\href {\doibase http://dx.doi.org/10.1016/0375-9601(91)90474-M}
  {\bibfield  {journal} {\bibinfo  {journal} {Physics Letters A}\ }\textbf
  {\bibinfo {volume} {155}},\ \bibinfo {pages} {229 } (\bibinfo {year}
  {1991})}\BibitemShut {NoStop}%
\bibitem [{\citenamefont {Korotkov}(1999)}]{CWLM1}%
  \BibitemOpen
  \bibfield  {author} {\bibinfo {author} {\bibfnamefont {A.~N.}\ \bibnamefont
  {Korotkov}},\ }\href@noop {} {\bibfield  {journal} {\bibinfo  {journal}
  {Phys. Rev. B}\ }\textbf {\bibinfo {volume} {60}},\ \bibinfo {pages} {5737}
  (\bibinfo {year} {1999})}\BibitemShut {NoStop}%
\bibitem [{\citenamefont {Jordan}\ and\ \citenamefont
  {Buttiker}(2005)}]{CWLM2}%
  \BibitemOpen
  \bibfield  {author} {\bibinfo {author} {\bibfnamefont {A.}~\bibnamefont
  {Jordan}}\ and\ \bibinfo {author} {\bibfnamefont {M.}~\bibnamefont
  {Buttiker}},\ }\href@noop {} {\bibfield  {journal} {\bibinfo  {journal}
  {Phys. Rev. Lett.}\ }\textbf {\bibinfo {volume} {95}},\ \bibinfo {pages}
  {220401} (\bibinfo {year} {2005})}\BibitemShut {NoStop}%
\bibitem [{\citenamefont {Jacobs}\ and\ \citenamefont {Steck}(2006)}]{CWLM25}%
  \BibitemOpen
  \bibfield  {author} {\bibinfo {author} {\bibfnamefont {K.}~\bibnamefont
  {Jacobs}}\ and\ \bibinfo {author} {\bibfnamefont {D.~A.}\ \bibnamefont
  {Steck}},\ }\href {\doibase 10.1080/00107510601101934} {\bibfield  {journal}
  {\bibinfo  {journal} {Contemporary Physics}\ }\textbf {\bibinfo {volume}
  {47}},\ \bibinfo {pages} {279} (\bibinfo {year} {2006})},\ \Eprint
  {http://arxiv.org/abs/http://dx.doi.org/10.1080/00107510601101934}
  {http://dx.doi.org/10.1080/00107510601101934} \BibitemShut {NoStop}%
\bibitem [{\citenamefont {Wei}\ and\ \citenamefont {Nazarov}(2008)}]{NazWei}%
  \BibitemOpen
  \bibfield  {author} {\bibinfo {author} {\bibfnamefont {H.}~\bibnamefont
  {Wei}}\ and\ \bibinfo {author} {\bibfnamefont {Y.~V.}\ \bibnamefont
  {Nazarov}},\ }\href {\doibase 10.1103/PhysRevB.78.045308} {\bibfield
  {journal} {\bibinfo  {journal} {Phys. Rev. B}\ }\textbf {\bibinfo {volume}
  {78}},\ \bibinfo {pages} {045308} (\bibinfo {year} {2008})}\BibitemShut
  {NoStop}%
\bibitem [{\citenamefont {Chantasri}\ and\ \citenamefont
  {Jordan}(2015)}]{CWLM3}%
  \BibitemOpen
  \bibfield  {author} {\bibinfo {author} {\bibfnamefont {A.}~\bibnamefont
  {Chantasri}}\ and\ \bibinfo {author} {\bibfnamefont {A.~N.}\ \bibnamefont
  {Jordan}},\ }\href@noop {} {\bibfield  {journal} {\bibinfo  {journal} {Phys.
  Rev. A}\ }\textbf {\bibinfo {volume} {92}},\ \bibinfo {pages} {032125}
  (\bibinfo {year} {2015})}\BibitemShut {NoStop}%
\bibitem [{\citenamefont {Chantasri}\ \emph {et~al.}(2013)\citenamefont
  {Chantasri}, \citenamefont {Dressel},\ and\ \citenamefont {Jordan}}]{CWLM4}%
  \BibitemOpen
  \bibfield  {author} {\bibinfo {author} {\bibfnamefont {A.}~\bibnamefont
  {Chantasri}}, \bibinfo {author} {\bibfnamefont {J.}~\bibnamefont {Dressel}},
  \ and\ \bibinfo {author} {\bibfnamefont {A.~N.}\ \bibnamefont {Jordan}},\
  }\href@noop {} {\bibfield  {journal} {\bibinfo  {journal} {Phys. Rev. A}\
  }\textbf {\bibinfo {volume} {88}},\ \bibinfo {pages} {042110} (\bibinfo
  {year} {2013})}\BibitemShut {NoStop}%
\bibitem [{\citenamefont {Hatridge}\ \emph {et~al.}(2013)\citenamefont
  {Hatridge}, \citenamefont {Shankar}, \citenamefont {Mirrahimi}, \citenamefont
  {Schackert}, \citenamefont {Geerlings}, \citenamefont {Brecht}, \citenamefont
  {Sliwa}, \citenamefont {Abdo}, \citenamefont {Frunzio}, \citenamefont
  {Girvin}, \citenamefont {Schoelkopf},\ and\ \citenamefont
  {Devoret}}]{Devoret}%
  \BibitemOpen
  \bibfield  {author} {\bibinfo {author} {\bibfnamefont {M.}~\bibnamefont
  {Hatridge}}, \bibinfo {author} {\bibfnamefont {S.}~\bibnamefont {Shankar}},
  \bibinfo {author} {\bibfnamefont {M.}~\bibnamefont {Mirrahimi}}, \bibinfo
  {author} {\bibfnamefont {F.}~\bibnamefont {Schackert}}, \bibinfo {author}
  {\bibfnamefont {K.}~\bibnamefont {Geerlings}}, \bibinfo {author}
  {\bibfnamefont {T.}~\bibnamefont {Brecht}}, \bibinfo {author} {\bibfnamefont
  {K.~M.}\ \bibnamefont {Sliwa}}, \bibinfo {author} {\bibfnamefont
  {B.}~\bibnamefont {Abdo}}, \bibinfo {author} {\bibfnamefont {L.}~\bibnamefont
  {Frunzio}}, \bibinfo {author} {\bibfnamefont {S.~M.}\ \bibnamefont {Girvin}},
  \bibinfo {author} {\bibfnamefont {R.~J.}\ \bibnamefont {Schoelkopf}}, \ and\
  \bibinfo {author} {\bibfnamefont {M.~H.}\ \bibnamefont {Devoret}},\ }\href
  {\doibase 10.1126/science.1226897} {\bibfield  {journal} {\bibinfo  {journal}
  {Science}\ }\textbf {\bibinfo {volume} {339}},\ \bibinfo {pages} {178}
  (\bibinfo {year} {2013})}\BibitemShut {NoStop}%
\bibitem [{\citenamefont {Murch}\ \emph {et~al.}(2013)\citenamefont {Murch},
  \citenamefont {Weber}, \citenamefont {Macklin},\ and\ \citenamefont
  {Siddiqi}}]{SiddiqiSingle}%
  \BibitemOpen
  \bibfield  {author} {\bibinfo {author} {\bibfnamefont {K.~W.}\ \bibnamefont
  {Murch}}, \bibinfo {author} {\bibfnamefont {S.~J.}\ \bibnamefont {Weber}},
  \bibinfo {author} {\bibfnamefont {C.}~\bibnamefont {Macklin}}, \ and\
  \bibinfo {author} {\bibfnamefont {I.}~\bibnamefont {Siddiqi}},\ }\href@noop
  {} {\bibfield  {journal} {\bibinfo  {journal} {Nature}\ }\textbf {\bibinfo
  {volume} {502}},\ \bibinfo {pages} {211} (\bibinfo {year}
  {2013})}\BibitemShut {NoStop}%
\bibitem [{\citenamefont {Roch}\ \emph {et~al.}(2014)\citenamefont {Roch},
  \citenamefont {Schwartz}, \citenamefont {Motzoi}, \citenamefont {Macklin},
  \citenamefont {Vijay}, \citenamefont {Eddins}, \citenamefont {Korotkov},
  \citenamefont {Whaley}, \citenamefont {Sarovar},\ and\ \citenamefont
  {Siddiqi}}]{SiddiqiEntanglement}%
  \BibitemOpen
  \bibfield  {author} {\bibinfo {author} {\bibfnamefont {N.}~\bibnamefont
  {Roch}}, \bibinfo {author} {\bibfnamefont {M.~E.}\ \bibnamefont {Schwartz}},
  \bibinfo {author} {\bibfnamefont {F.}~\bibnamefont {Motzoi}}, \bibinfo
  {author} {\bibfnamefont {C.}~\bibnamefont {Macklin}}, \bibinfo {author}
  {\bibfnamefont {R.}~\bibnamefont {Vijay}}, \bibinfo {author} {\bibfnamefont
  {A.~W.}\ \bibnamefont {Eddins}}, \bibinfo {author} {\bibfnamefont {A.~N.}\
  \bibnamefont {Korotkov}}, \bibinfo {author} {\bibfnamefont {K.~B.}\
  \bibnamefont {Whaley}}, \bibinfo {author} {\bibfnamefont {M.}~\bibnamefont
  {Sarovar}}, \ and\ \bibinfo {author} {\bibfnamefont {I.}~\bibnamefont
  {Siddiqi}},\ }\href {\doibase 10.1103/PhysRevLett.112.170501} {\bibfield
  {journal} {\bibinfo  {journal} {Phys. Rev. Lett.}\ }\textbf {\bibinfo
  {volume} {112}},\ \bibinfo {pages} {170501} (\bibinfo {year}
  {2014})}\BibitemShut {NoStop}%
\bibitem [{\citenamefont {Campagne-Ibarcq}\ \emph {et~al.}(2014)\citenamefont
  {Campagne-Ibarcq}, \citenamefont {Bretheau}, \citenamefont {Flurin},
  \citenamefont {Auff\`eves}, \citenamefont {Mallet},\ and\ \citenamefont
  {Huard}}]{Huard}%
  \BibitemOpen
  \bibfield  {author} {\bibinfo {author} {\bibfnamefont {P.}~\bibnamefont
  {Campagne-Ibarcq}}, \bibinfo {author} {\bibfnamefont {L.}~\bibnamefont
  {Bretheau}}, \bibinfo {author} {\bibfnamefont {E.}~\bibnamefont {Flurin}},
  \bibinfo {author} {\bibfnamefont {A.}~\bibnamefont {Auff\`eves}}, \bibinfo
  {author} {\bibfnamefont {F.}~\bibnamefont {Mallet}}, \ and\ \bibinfo {author}
  {\bibfnamefont {B.}~\bibnamefont {Huard}},\ }\href {\doibase
  10.1103/PhysRevLett.112.180402} {\bibfield  {journal} {\bibinfo  {journal}
  {Phys. Rev. Lett.}\ }\textbf {\bibinfo {volume} {112}},\ \bibinfo {pages}
  {180402} (\bibinfo {year} {2014})}\BibitemShut {NoStop}%
\bibitem [{\citenamefont {Groen}\ \emph {et~al.}(2013)\citenamefont {Groen},
  \citenamefont {Rist\`e}, \citenamefont {Tornberg}, \citenamefont {Cramer},
  \citenamefont {de~Groot}, \citenamefont {Picot}, \citenamefont {Johansson},\
  and\ \citenamefont {DiCarlo}}]{DiCarlo}%
  \BibitemOpen
  \bibfield  {author} {\bibinfo {author} {\bibfnamefont {J.~P.}\ \bibnamefont
  {Groen}}, \bibinfo {author} {\bibfnamefont {D.}~\bibnamefont {Rist\`e}},
  \bibinfo {author} {\bibfnamefont {L.}~\bibnamefont {Tornberg}}, \bibinfo
  {author} {\bibfnamefont {J.}~\bibnamefont {Cramer}}, \bibinfo {author}
  {\bibfnamefont {P.~C.}\ \bibnamefont {de~Groot}}, \bibinfo {author}
  {\bibfnamefont {T.}~\bibnamefont {Picot}}, \bibinfo {author} {\bibfnamefont
  {G.}~\bibnamefont {Johansson}}, \ and\ \bibinfo {author} {\bibfnamefont
  {L.}~\bibnamefont {DiCarlo}},\ }\href {\doibase
  10.1103/PhysRevLett.111.090506} {\bibfield  {journal} {\bibinfo  {journal}
  {Phys. Rev. Lett.}\ }\textbf {\bibinfo {volume} {111}},\ \bibinfo {pages}
  {090506} (\bibinfo {year} {2013})}\BibitemShut {NoStop}%
\bibitem [{\citenamefont {Weber}\ \emph {et~al.}(2014)\citenamefont {Weber},
  \citenamefont {Chantasri}, \citenamefont {Dressel}, \citenamefont {Jordan},
  \citenamefont {Murch},\ and\ \citenamefont {Siddiqi}}]{SiddiqiMapping}%
  \BibitemOpen
  \bibfield  {author} {\bibinfo {author} {\bibfnamefont {S.~J.}\ \bibnamefont
  {Weber}}, \bibinfo {author} {\bibfnamefont {A.}~\bibnamefont {Chantasri}},
  \bibinfo {author} {\bibfnamefont {J.}~\bibnamefont {Dressel}}, \bibinfo
  {author} {\bibfnamefont {A.~N.}\ \bibnamefont {Jordan}}, \bibinfo {author}
  {\bibfnamefont {K.~W.}\ \bibnamefont {Murch}}, \ and\ \bibinfo {author}
  {\bibfnamefont {I.}~\bibnamefont {Siddiqi}},\ }\href@noop {} {\bibfield
  {journal} {\bibinfo  {journal} {Nature}\ }\textbf {\bibinfo {volume} {511}},\
  \bibinfo {pages} {570} (\bibinfo {year} {2014})}\BibitemShut {NoStop}%
\bibitem [{\citenamefont {Tan}\ \emph {et~al.}(2015)\citenamefont {Tan},
  \citenamefont {Weber}, \citenamefont {Siddiqi}, \citenamefont {M\o{}lmer},\
  and\ \citenamefont {Murch}}]{SiddiqiMolmer}%
  \BibitemOpen
  \bibfield  {author} {\bibinfo {author} {\bibfnamefont {D.}~\bibnamefont
  {Tan}}, \bibinfo {author} {\bibfnamefont {S.~J.}\ \bibnamefont {Weber}},
  \bibinfo {author} {\bibfnamefont {I.}~\bibnamefont {Siddiqi}}, \bibinfo
  {author} {\bibfnamefont {K.}~\bibnamefont {M\o{}lmer}}, \ and\ \bibinfo
  {author} {\bibfnamefont {K.~W.}\ \bibnamefont {Murch}},\ }\href {\doibase
  10.1103/PhysRevLett.114.090403} {\bibfield  {journal} {\bibinfo  {journal}
  {Phys. Rev. Lett.}\ }\textbf {\bibinfo {volume} {114}},\ \bibinfo {pages}
  {090403} (\bibinfo {year} {2015})}\BibitemShut {NoStop}%
\bibitem [{\citenamefont {Wiseman}(2002)}]{WisemanWeakValues}%
  \BibitemOpen
  \bibfield  {author} {\bibinfo {author} {\bibfnamefont {H.~M.}\ \bibnamefont
  {Wiseman}},\ }\href {\doibase 10.1103/PhysRevA.65.032111} {\bibfield
  {journal} {\bibinfo  {journal} {Phys. Rev. A}\ }\textbf {\bibinfo {volume}
  {65}},\ \bibinfo {pages} {032111} (\bibinfo {year} {2002})}\BibitemShut
  {NoStop}%
\bibitem [{\citenamefont {Franquet}\ and\ \citenamefont
  {Nazarov}(2017{\natexlab{a}})}]{FranquetNaz}%
  \BibitemOpen
  \bibfield  {author} {\bibinfo {author} {\bibfnamefont {A.}~\bibnamefont
  {Franquet}}\ and\ \bibinfo {author} {\bibfnamefont {Y.~V.}\ \bibnamefont
  {Nazarov}},\ }\href {\doibase 10.1103/PhysRevB.95.085427} {\bibfield
  {journal} {\bibinfo  {journal} {Phys. Rev. B}\ }\textbf {\bibinfo {volume}
  {95}},\ \bibinfo {pages} {085427} (\bibinfo {year}
  {2017}{\natexlab{a}})}\BibitemShut {NoStop}%
\bibitem [{\citenamefont {Franquet}\ and\ \citenamefont
  {Nazarov}(2017{\natexlab{b}})}]{FranquetNaz2}%
  \BibitemOpen
  \bibfield  {author} {\bibinfo {author} {\bibfnamefont {A.}~\bibnamefont
  {Franquet}}\ and\ \bibinfo {author} {\bibfnamefont {Y.~V.}\ \bibnamefont
  {Nazarov}},\ }\href@noop {} {\bibfield  {journal} {\bibinfo  {journal}
  {arXiv:1708.05662 [quant-ph]}\ } (\bibinfo {year}
  {2017}{\natexlab{b}})}\BibitemShut {NoStop}%
\bibitem [{\citenamefont {Vijay}\ \emph {et~al.}(2012)\citenamefont {Vijay},
  \citenamefont {Macklin}, \citenamefont {Slichter}, \citenamefont {Weber},
  \citenamefont {Murch}, \citenamefont {Naik}, \citenamefont {Korotkov},\ and\
  \citenamefont {Siddiqi}}]{SiddiqiFeedback}%
  \BibitemOpen
  \bibfield  {author} {\bibinfo {author} {\bibfnamefont {R.}~\bibnamefont
  {Vijay}}, \bibinfo {author} {\bibfnamefont {C.}~\bibnamefont {Macklin}},
  \bibinfo {author} {\bibfnamefont {D.~H.}\ \bibnamefont {Slichter}}, \bibinfo
  {author} {\bibfnamefont {S.~J.}\ \bibnamefont {Weber}}, \bibinfo {author}
  {\bibfnamefont {K.~W.}\ \bibnamefont {Murch}}, \bibinfo {author}
  {\bibfnamefont {R.}~\bibnamefont {Naik}}, \bibinfo {author} {\bibfnamefont
  {A.~N.}\ \bibnamefont {Korotkov}}, \ and\ \bibinfo {author} {\bibfnamefont
  {I.}~\bibnamefont {Siddiqi}},\ }\href@noop {} {\bibfield  {journal} {\bibinfo
   {journal} {Nature}\ }\textbf {\bibinfo {volume} {490}},\ \bibinfo {pages}
  {77} (\bibinfo {year} {2012})}\BibitemShut {NoStop}%
\bibitem [{\citenamefont {de~Lange}\ \emph {et~al.}(2014)\citenamefont
  {de~Lange}, \citenamefont {Rist\`e}, \citenamefont {Tiggelman}, \citenamefont
  {Eichler}, \citenamefont {Tornberg}, \citenamefont {Johansson}, \citenamefont
  {Wallraff}, \citenamefont {Schouten},\ and\ \citenamefont
  {DiCarlo}}]{AnalogFeedback}%
  \BibitemOpen
  \bibfield  {author} {\bibinfo {author} {\bibfnamefont {G.}~\bibnamefont
  {de~Lange}}, \bibinfo {author} {\bibfnamefont {D.}~\bibnamefont {Rist\`e}},
  \bibinfo {author} {\bibfnamefont {M.~J.}\ \bibnamefont {Tiggelman}}, \bibinfo
  {author} {\bibfnamefont {C.}~\bibnamefont {Eichler}}, \bibinfo {author}
  {\bibfnamefont {L.}~\bibnamefont {Tornberg}}, \bibinfo {author}
  {\bibfnamefont {G.}~\bibnamefont {Johansson}}, \bibinfo {author}
  {\bibfnamefont {A.}~\bibnamefont {Wallraff}}, \bibinfo {author}
  {\bibfnamefont {R.~N.}\ \bibnamefont {Schouten}}, \ and\ \bibinfo {author}
  {\bibfnamefont {L.}~\bibnamefont {DiCarlo}},\ }\href {\doibase
  10.1103/PhysRevLett.112.080501} {\bibfield  {journal} {\bibinfo  {journal}
  {Phys. Rev. Lett.}\ }\textbf {\bibinfo {volume} {112}},\ \bibinfo {pages}
  {080501} (\bibinfo {year} {2014})}\BibitemShut {NoStop}%
\bibitem [{\citenamefont {Greplova}\ \emph {et~al.}(2016)\citenamefont
  {Greplova}, \citenamefont {M\o{}lmer},\ and\ \citenamefont
  {Andersen}}]{simulation1}%
  \BibitemOpen
  \bibfield  {author} {\bibinfo {author} {\bibfnamefont {E.}~\bibnamefont
  {Greplova}}, \bibinfo {author} {\bibfnamefont {K.}~\bibnamefont {M\o{}lmer}},
  \ and\ \bibinfo {author} {\bibfnamefont {C.~K.}\ \bibnamefont {Andersen}},\
  }\href {\doibase 10.1103/PhysRevA.94.042334} {\bibfield  {journal} {\bibinfo
  {journal} {Phys. Rev. A}\ }\textbf {\bibinfo {volume} {94}},\ \bibinfo
  {pages} {042334} (\bibinfo {year} {2016})}\BibitemShut {NoStop}%
\bibitem [{\citenamefont {Jack}\ and\ \citenamefont
  {Collett}(2000)}]{simulation2}%
  \BibitemOpen
  \bibfield  {author} {\bibinfo {author} {\bibfnamefont {M.~W.}\ \bibnamefont
  {Jack}}\ and\ \bibinfo {author} {\bibfnamefont {M.~J.}\ \bibnamefont
  {Collett}},\ }\href {\doibase 10.1103/PhysRevA.61.062106} {\bibfield
  {journal} {\bibinfo  {journal} {Phys. Rev. A}\ }\textbf {\bibinfo {volume}
  {61}},\ \bibinfo {pages} {062106} (\bibinfo {year} {2000})}\BibitemShut
  {NoStop}%
\bibitem [{\citenamefont {Ralph}\ \emph {et~al.}(2006)\citenamefont {Ralph},
  \citenamefont {Bartlett}, \citenamefont {O'Brien}, \citenamefont {Pryde},\
  and\ \citenamefont {Wiseman}}]{nondemolition}%
  \BibitemOpen
  \bibfield  {author} {\bibinfo {author} {\bibfnamefont {T.~C.}\ \bibnamefont
  {Ralph}}, \bibinfo {author} {\bibfnamefont {S.~D.}\ \bibnamefont {Bartlett}},
  \bibinfo {author} {\bibfnamefont {J.~L.}\ \bibnamefont {O'Brien}}, \bibinfo
  {author} {\bibfnamefont {G.~J.}\ \bibnamefont {Pryde}}, \ and\ \bibinfo
  {author} {\bibfnamefont {H.~M.}\ \bibnamefont {Wiseman}},\ }\href {\doibase
  10.1103/PhysRevA.73.012113} {\bibfield  {journal} {\bibinfo  {journal} {Phys.
  Rev. A}\ }\textbf {\bibinfo {volume} {73}},\ \bibinfo {pages} {012113}
  (\bibinfo {year} {2006})}\BibitemShut {NoStop}%
\bibitem [{\citenamefont {Clerk}\ \emph {et~al.}(2010)\citenamefont {Clerk},
  \citenamefont {Devoret}, \citenamefont {Girvin}, \citenamefont {Marquardt},\
  and\ \citenamefont {Schoelkopf}}]{QNoise}%
  \BibitemOpen
  \bibfield  {author} {\bibinfo {author} {\bibfnamefont {A.~A.}\ \bibnamefont
  {Clerk}}, \bibinfo {author} {\bibfnamefont {M.~H.}\ \bibnamefont {Devoret}},
  \bibinfo {author} {\bibfnamefont {S.~M.}\ \bibnamefont {Girvin}}, \bibinfo
  {author} {\bibfnamefont {F.}~\bibnamefont {Marquardt}}, \ and\ \bibinfo
  {author} {\bibfnamefont {R.~J.}\ \bibnamefont {Schoelkopf}},\ }\href
  {\doibase 10.1103/RevModPhys.82.1155} {\bibfield  {journal} {\bibinfo
  {journal} {Rev. Mod. Phys.}\ }\textbf {\bibinfo {volume} {82}},\ \bibinfo
  {pages} {1155} (\bibinfo {year} {2010})}\BibitemShut {NoStop}%
\end{thebibliography}%
\end{document}